\documentclass[tighten,times]{aa}  
\usepackage{graphicx}
\usepackage{txfonts}
\usepackage{natbib}
\usepackage{xcolor}
\usepackage{hyperref}

\begin{document} 

   \title{On Collision Course: The Nature of the Binary Star Cluster NGC\,2006/SL538}


   \author{Marcelo~D.~Mora\inst{1}
          \and
          Thomas~H.~Puzia\inst{1}
          \and
          Julio Chanam\'e\inst{1}
          }

   \institute{Institute of Astrophysics, Pontificia Universidad Cat\'olica de Chile, Av. Vicu\~na Mackenna 4860, Macul 7820436, Santiago, Chile, \\
    \email{mmora@astro.puc.cl}}

   \date{Received \today; accepted tomorrow}

 
  \abstract
   {The LMC is known to be the host of a rich variety of star clusters of all ages. A large number of them is seen in close projected proximity.~Ages have been derived for few of them showing differences up to few million years, hinting at being binary star clusters. However, final confirmation needs to be done through spectroscopy measurements and dynamical analysis. }
  {In the present work we focus on one of these LMC cluster pairs (\object{NGC\,2006} - \object{SL\,538}) and aim to determine whether the star cluster pair is a bound entity and, therefore, a binary star cluster or a chance alignment.}
   {Using the MIKE echelle high-resolution spectrograph on the 6.5-meter Magellan-II Clay telescope at Las Campanas Observatory we have acquired integrated-light spectra of the two clusters, measuring their radial velocities with individual absorption features and cross-correlation of each spectrum with  a stellar spectral library.}
   {We have measured radial velocities by two methods: The first one by direct line profile measurement yields $v_r\!=\!300.3\pm5\pm6$ km s$^{-1}$ for NGC\,2006 and $v_r\!=\!310.2\pm4\pm6$ km s$^{-1}$ for SL\,538.~The second one is derived by comparing observed spectra with synthetic bootstrapped spectra yielding $v_r\!=\!311.0\pm0.6$ km s$^{-1}$ for NGC\,2006 and $v_r\!=\!309.4\pm0.5$ km s$^{-1}$ for SL\,538.~Finally when spectra are directly compared, we find a $\Delta v\!=\!1.08\pm0.47$ km s$^{-1}$.~Full-spectrum SED fits reveal that the stellar population ages of both clusters lie in the range $13\!-\!21$ Myr with a metallicity of $Z\!=\!0.008$. We find indications for differences in the chemical abundance patterns as revealed by the helium absorption lines between the two clusters. The dynamical analysis of the system shows that the two clusters are likely to merge within the next $\sim\!150$ Myr to form a star cluster with a stellar mass of $\sim\!10^4\,M_\odot$}
   {The NGC\,2006--SL\,538 cluster pair shows radial velocities, stellar population and dynamical parameters consistent with a gravitational bound entity and, considering that the velocity dispersion of the stars in LMC is $\lesssim\!20$ km/s, we reject them as a chance alignment. We conclude that this is a genuine binary cluster pair, and we propose that their differences in ages and stellar population chemistry is most likely due to variances in their chemical enrichment history within their environment. We suggest that their formation may have taken place in a loosely bound star-formation complex which saw initial fragmentation but then had its clusters become a gravitationally bound pair by tidal capture.}

	\keywords{Galaxies: star clusters: individual: NGC2006, SL538 -- Galaxies: individual: LMC}
 
   \maketitle
%

\section{Introduction}
In the context of star cluster formation, it is known that star clusters form in large fractal entities where interactions between sub-clusters of various masses are likely.~These interactions are short and violent processes that happen at the youngest stages of star cluster evolution and last for a few to a few tens of millions of years \citep{Sugimoto:1989qy, Yu:2017}.~Such processes may lead to star cluster disruption or star cluster merging \cite[e.g.][]{Kroupa1998,Fellhauer:2005kx}. A snapshot of this short-living stage may be observed as two star clusters in close projected proximity to each other on the sky (i.e.~star cluster pairs). Such pairs may then be considered binary star clusters if their relative velocities are consistent with being gravitationally bound systems.

The origin of binary clusters, as proposed by \citep{Fujimoto:1997uq}, is a parent gas cloud that fragments in at least two separate sub-clumps, yielding two or more bound star clusters of virtually identical ages and abundances. However, it is also possible to form binary star clusters by tidal capture \citep{Leon:1999sf}, where clusters born in different clouds (thus likely having different ages and abundances) later become a bound system due to a close encounter and subsequent loss of angular momentum. If these sub-clusters are gravitationally bound, they will eventually merge and form a more massive star cluster, mixing the constituent stellar populations of the sub-clusters. Such processes might be one of the possible scenarios that may lead to globular clusters with multiple stellar populations \citep{Gratton:2012jk}, which is likely to happen in dwarf galaxies where the stellar velocity dispersion within the galaxy is comparable to that within the globular clusters, e.g. Sagittarius dwarf \citep{Bellazzini:2008fp} and the Magellanic clouds \citep[e.g.][]{Bhatia:1988eu}. 
  
\begin{figure*}[t!]
\centering
\includegraphics[width=0.95\textwidth,angle=0]{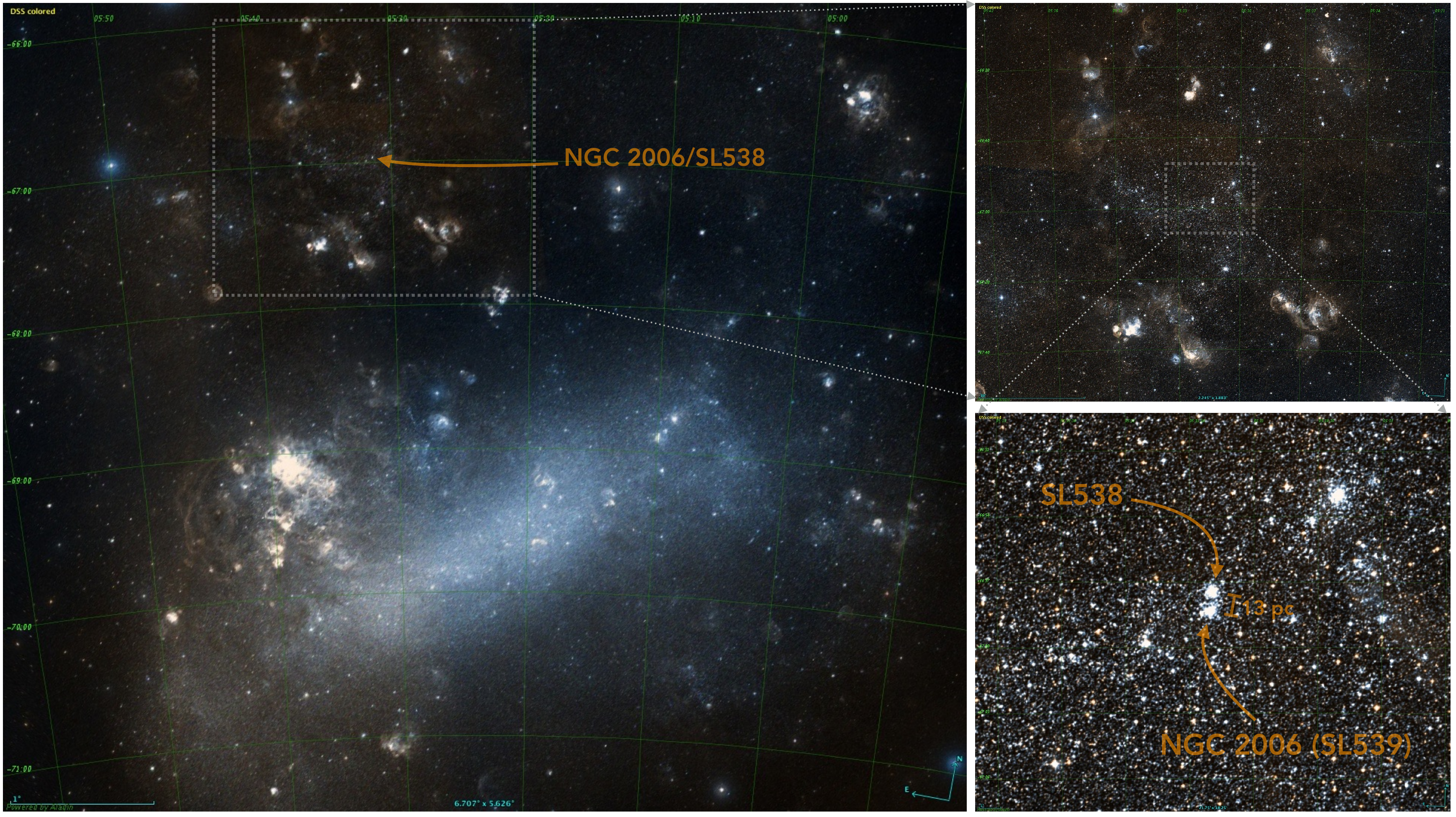}
\caption{Overview of the Large Magellanic Cloud (LMC) and the location, environment and morphology of the NGC\,2006 (SL539) and SL538 star cluster pair. The figures are based on DSS colored images. The left panel is an overview of the LMC with the location of NGC\,2006/SL539 indicated by the arrow. The dashed square shows the outline of the zoom-in image in the top right panel, which illustrates the environment of the double star cluster. The bottom right panel shows the star cluster pair, separated by $\sim\!13$\,pc, within the stellar field of the immediate environment.}
\label{Picture_pair}
\end{figure*}

The Magellanic clouds host several star clusters pairs whose stellar populations have been analyzed photometrically by means of color-magnitude diagrams \citep[e.g.][]{Bhatia:1988eu,Hatzidimitriou:1990yq, Bhatia:1991rt, Pietrzynski:1999zr}. For example, \citet{Dieball:2000ly} concluded that the pair NGC\,1971 \& NGC\,1972 have similar ages ($40-70$ Myr) indicating that both clusters may have been formed in the same giant molecular cloud. However, they could not establish whether the clusters are physically interacting or not. They also found that NGC\,1894 \& SL\,341 show similar ages ($55\pm5$ Myr), concluding that the formation of both clusters from the same giant molecular cloud is likely. Using Str\"omgren CCD photometry \citet{Hilker:1995ul} found that NGC\,2136 \& NGC\,2137 have common metallicities ($-0.55$ dex), also their color magnitude diagrams lie along the same isochrone and, therefore, the ages of the two star clusters cannot be distinguished, concluding that the simultaneous formation is likely. Later \citet{Dirsch:2000bh} derived an age of $100\pm25$ Myr for this star cluster pair. 
\citet{Dieball:2000ly} combined stellar photometry of SL\,353 and SL\,349 with spectroscopic measurements for 22 stars on both clusters and found an age of $500\pm100$ Myr for both clusters and radial velocities of $v=274 \pm 4$ and $v=279 \pm 19$ km/s for SL\,353 and SL\,349, respectively. They argue that the small velocity difference and similar ages are consistent with a gravitationally bound star cluster pair. 
Recently, \citep{Mucciarelli:2012fk} found radial velocities $v=271.5\pm0.4$ and $270.6\pm0.5$ km/s for NGC\,2136 and NGC\,2137, and indistinguishable abundance patterns [Fe/H] =$-0.40 \pm 0.01$ and $-0.39 \pm 0.01$ dex, respectively. Based on the numerical simulations of this system performed by \citet{Portegies-Zwart:2007rw}, the orbital parameters indicate that a merger is expected on a timescale comparable to the orbital period of these clusters.

The star cluster pair NGC\,2006 (SL539)/SL538 (see Fig.~\ref{Picture_pair}) is located in the outskirts of the Large Magellanic Cloud (LMC) and is one of the many such pairs detected in the LMC \citep{Dieball:2002zr}. Both clusters are seen at a projected distance of 13.3 pc \citep{Dieball:1998rm}, and are located in the supergiant shell LMC\,4, at the northwestern part of the OB association LH\,77.~Several authors have measured ages of $22.5\pm2.5$ Myr \citep{Dieball:1998rm} and $25\pm3$ Myr \citep{Kumar:2008rw} for NGC\,2006, while SL538 was found to have ages of $18\pm2$ Myr \citep{Dieball:1998rm} and $20\pm2$ Myr \citep{Kumar:2008rw}. Both clusters were reported as binary clusters by \citet{Kontizas:1993yg} through the analysis of the cores of the clusters using low-resolution prism spectra and integrated IUE spectra.~The LMC is not the only galaxy hosting confirmed and likely binary clusters. For example, \citep{De-Silva:2015fj} reported the open cluster pair NGC~5617/Trumpet~22 as primordial binary cluster pair in the Milky Way, while likely binary clusters have also been found in NGC\,5128 \citep{Minniti:2004ty}. Therefore, binary star clusters are being detected in more galaxies with recent or ongoing star-formation activity. Whether they are a common entity or not needs to be investigated in greater detail.

For a cluster pair to be considered a binary star cluster, it must fulfill the following conditions: Both clusters must be seen as a star cluster pair (i.e.~in close projected proximity).~Clusters may share similar/different ages and abundances (under the assumption of a common/different original cloud), and a difference in radial velocity of few km s$^{-1}$. Under these criteria, if a star cluster pair like NGC\,2006 (SL539)/SL538 in the LMC is bound (assuming masses of the order $10^{4}\,$M$_{\sun}$) the expected orbital velocity correspond indeed to a few km s$^{-1}$, which in low-resolution spectra will be seen as identical velocities \citep[e.g.][]{Kontizas:1993yg}. For younger stellar populations, and by inference for younger star clusters, the velocity dispersion is lower at about $\sigma\approx20$ km s$^{-1}$ \citep[e.g.][]{2002AJ....124.2639V}. On the other hand, if the star cluster pair is a chance alignment the radial velocity difference is expected to be consistent with the stellar field velocity dispersion of the younger stellar population in the LMC \citep[i.e.~likely smaller than 20 km s$^{-1}$; see][]{2002AJ....124.2639V}.
 Thus, in order to move one step towards the confirmation of binarity (or its rejection) high-resolution spectroscopy is required to accurately measure radial velocities with km s$^{-1}$ accuracy and distinguish between these two scenarios. This was done by \citep{Mucciarelli:2012fk} in the case of the star cluster pair NGC\,2136/2137 which led to the confirmation of its binary status.~In this work we present the kinematic analysis of the cluster pair NGC\,2006 (SL539)/SL538 using high-resolution spectra with the aim of constraining whether this pair is a bound system. Our paper is structured as follows: In Section~\ref{ln:obs} we describe the observations, while in Section~\ref{ln:datared} we describe the data reduction. Section~\ref{ln:anls} describes our analysis and finally in Section~\ref{ln:disc} we discuss the results and present our conclusions.
  
\begin{figure*}[!ht]
\centering
\includegraphics[width=0.95\textwidth,angle=0]{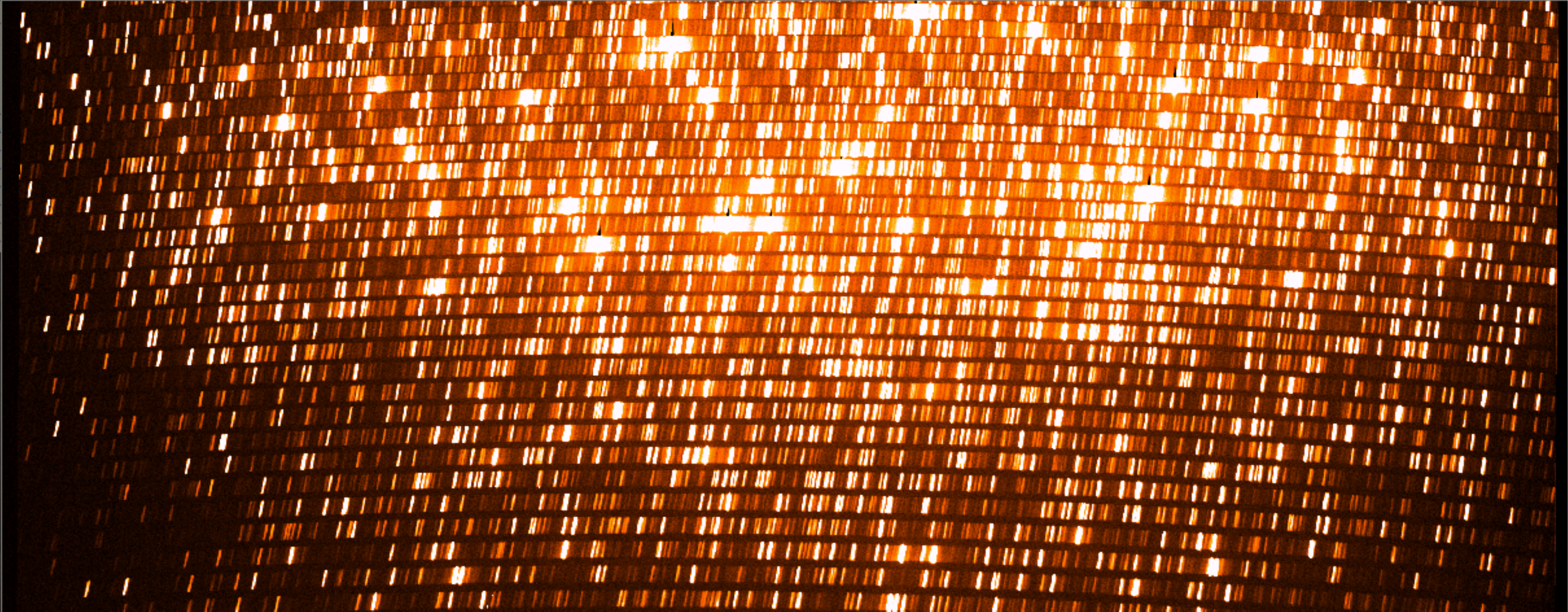}
\caption{Example of an arc frame from the MIKE spectrograph.~Note that individual orders are heavily curved, as typical for echelle spectra, in addition to a substantial tilt of the emission lines within each order that changes as a function of wavelength.}
\label{Curved}
\end{figure*}

\section{Observations}
\label{ln:obs}
Spectra were acquired during the night of December 10th, 2013 with the Magellan Inamori Kyocera Echelle (MIKE) spectrograph mounted on the Magellan-II Clay telescope, located at Las Campanas Observatory in Chile. The instrument setup was using the $0.7\arcsec\times5.0\arcsec$ slit and the CCD readout was binned in $2\times2$ pixels, which yields a spectral resolution of $\sim\!30,000$ in the blue part of the spectrum.~The wavelength coverage spans 3600 to 9000\,\AA.~The night was photometric with a seeing of $\sim\!0.6\arcsec$. For  each cluster we pointed toward the central part and we applied  a random jitter inside the central cluster area without closing the shutter with the aim of obtaining a representative sample of the stellar population of each cluster. Three exposures of 900 second were taken for each cluster and one exposure of 900 sec in a nearby empty field with the aim of deriving and subtracting the average integrated-light spectrum of the sky and background.

\section{Data reduction}
\label{ln:datared}
High-resolution echelle spectra require careful data reduction in order to obtain the best radial velocity results of the star cluster pair.~One of the critical issues regarding echelle data reduction is the correction of the individual orders which are typically heavily curved.~In the case of our MIKE spectrograph, the instrument optics introduces an additional tilt of the wavelength solution within one order, as can be seen in an arc frame shown in Figure~\ref{Curved}. Our integrated-light spectra require the use of the entire (fully illuminated) slit and not only the trace of a point source, as it is traditional done in the majority of spectroscopic observations. Accurate correction of the differential wavelength calibration of every pixel in Figure~\ref{Curved} is necessary before collapsing the order into a 1D spectrum. Without such a correction absorption and emission features would be washed out and spectral resolution would be lost, and in the most extreme cases, the features might appear double-peaked. In the following we describe a custom recipe that we developed to apply a pixel-by-pixel wavelength calibration to the individual echelle orders. We make this recipe publicly available\footnote{Available at \href{https://github.com/mmorage/LCO_ECHELLE}{https://github.com/mmorage/LCO\_ECHELLE}} and encourage comments and suggestion for improvement from the community.

The correction procedure assumes that the science frames are bias and flat-field corrected. We note that it is critical to acquire well-illuminated flat-field frames, even if it saturates some of the orders, however, not to the point of ``bleeding". Such flats are used to straighten out the individual orders. The idea is to use the order illumination patterns to map the optics curvature, which is then used to correct individual orders and transform them into rectified orders. Poorly illuminated orders can be successfully corrected after initially applying a smoothing function, which can enhance the order-edge detectability at the expense of cutting a fraction of the spatial extent of the sky. Before appliying our recipe, bias and flatfield corrected science frames were cosmic cleaned using the La-cosmic  \citep{van-Dokkum:2001fk} routine. 

\subsection{Order edge detection and mapping}
\begin{figure}[t]
\centering
\includegraphics[width=\columnwidth,angle=0]{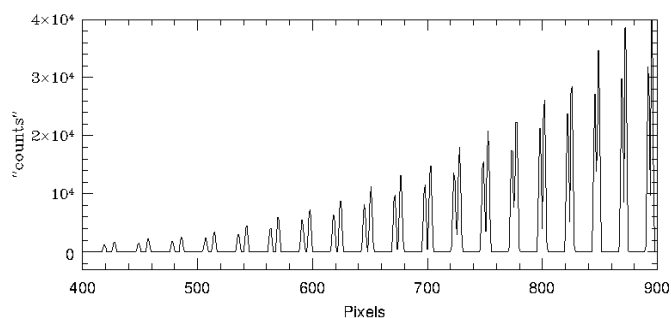}
\caption{Cut along the y-axis of the "pure-edge" image. Edge borders are indicated by positive count values and show a characteristic change in order width as a function of y-coordinate pixels, which corresponds to the spatial direction along the slit.}
\label{edges}
\end{figure}

The first step uses the flat-field frames to identify and characterize the edges of the individual echelle orders. Poorly illuminated orders are smoothed along the x-axis (roughly corresponding to the dispersion axis) to obtain better defined edges. Once the order edges are sufficiently well defined, we apply a {\sc Sobel} edge detection algorithm to the image along the y-axis using the {\sc ndimage.sobel SciPy} task. This task  creates a new image with highlighted order edges. This highlighted-edge image is then subtracted from the original flat-field frame creating a "pure-edge" image, which contains only information regarding the position of the edges. Finally, an (adjustable) minimum threshold of 500 "counts" is used to avoid inter-order features. The final product is a pure-edge image which corresponds to the science frame with zeros in the order and inter-order regions, except for the detected edges. Figure~\ref{edges} illustrates a cut along the y-axis through the pure-edge image.

To map each order we use a simple Python script that extracts the order regions that lie in between the previously detected order edges and stores them in a look-up coordinate table. For each order the output is visually verified on the image to avoid spurious detections typically at the extreme ends along the dispersion direction in each order. This procedure works well for well illuminated orders, but it requires human interaction for order regions where the flat-field frame is poorly illuminated. Nevertheless, since the important information is stored in the tables with the edge coordinates, it is still possible to cut, straighten or extrapolate the orders manually, if necessary.

\subsection{Order rectification}
The next step concerns the straightening of individual orders. For this purpose, we use the mapped edge coordinates of each order and use the {\sc IRAF} task \emph{geomap} to derive a transformation from the curved into the rectified frame. Since \emph{geomap} requires the target coordinates to be specified, we adopt a uniform slit length of 26 pixels (5\arcsec on the sky) in the target frame in order to preserve a uniform spatial scaling of the slit in each order. Since this is an interactive task, outliers can be easily corrected by deleting the corresponding points that are outside the regular shape of the curved order. We use a {\sc spline2} function with an order of 3 for the rectification. Finally, the {\sc Pyraf} task \emph{gregister} is used to execute the transformation with the option of flux conservation set to yes. An example of the final outcome from this process can be seen in the top panel of Figure~\ref{order_corrected} where we show a straighten order, however, still with a tilted wavelength solution.

\subsection{Wavelength calibration and sky subtraction}
\begin{figure}[b]
\centering
\includegraphics[width=\columnwidth,angle=0]{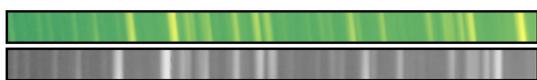} 
\caption{Example of the order rectification and wavelength calibration correction. ({\it Top panel}): rectified order cutout of an arc frame, not wavelength calibrated. ({\it Bottom panel}): wavelength calibrated arc frame.}
\label{order_corrected}
\end{figure}

Since the wavelength solution in the rectified frame is still tilted, i.e.~the arc emission lines have not one the same x-coordinate at any y-coordinate, a proper wavelength calibration of each pixel in the rectified order needs to be applied. This was done using the {\sc Pyraf} task \emph{identify} in a classical way, for each pixel row of each order. Typical r.m.s.~values for the wavelength solutions were of the order of 0.4\,\AA\ and the final dispersion correction was applied using the \emph{dispcor} task. A rectified and wavelength calibrated frame is illustrated in the bottom panel of Figure~\ref{order_corrected}.

\subsection{Science spectra and sky subtraction}
For each order, we summed all rows of the rectified and wavelength calibrated frames into a single object spectrum, i.e.~we collapsed the 2D spectra into a combined 1D spectrum.~Since three exposures were acquired for each cluster, the final step was the addition of these three collapsed spectra yielding the final science spectrum for each cluster.

Background subtraction was performed using the sky observations obtained with the corresponding science frames. For each order, we created an average combined master sky spectrum which was identically reduced as the science frames and accordingly scaled. These 1D sky spectra were then subtracted from the corresponding 1D science spectra.

\section{Analysis}
\label{ln:anls}
Prior to the kinematic analysis, we explore the individual orders and find that the most prominent spectral features are detected on the blue-side detector of MIKE, as it is expected for a young stellar population. Since we obtain a significantly higher signal-to-noise ratio (S/N) on the blue-side than on the red-side detector, we focus our subsequent analysis on the blue side of MIKE which samples the wavelength region of about $3600-5000$ \AA\ at a resolution of $R\approx28\,000$ with a sampling of $\sim\!0.04$ \AA/pix.

\begin{figure*}[!ht]
\centering
\includegraphics[width=\textwidth,angle=0]{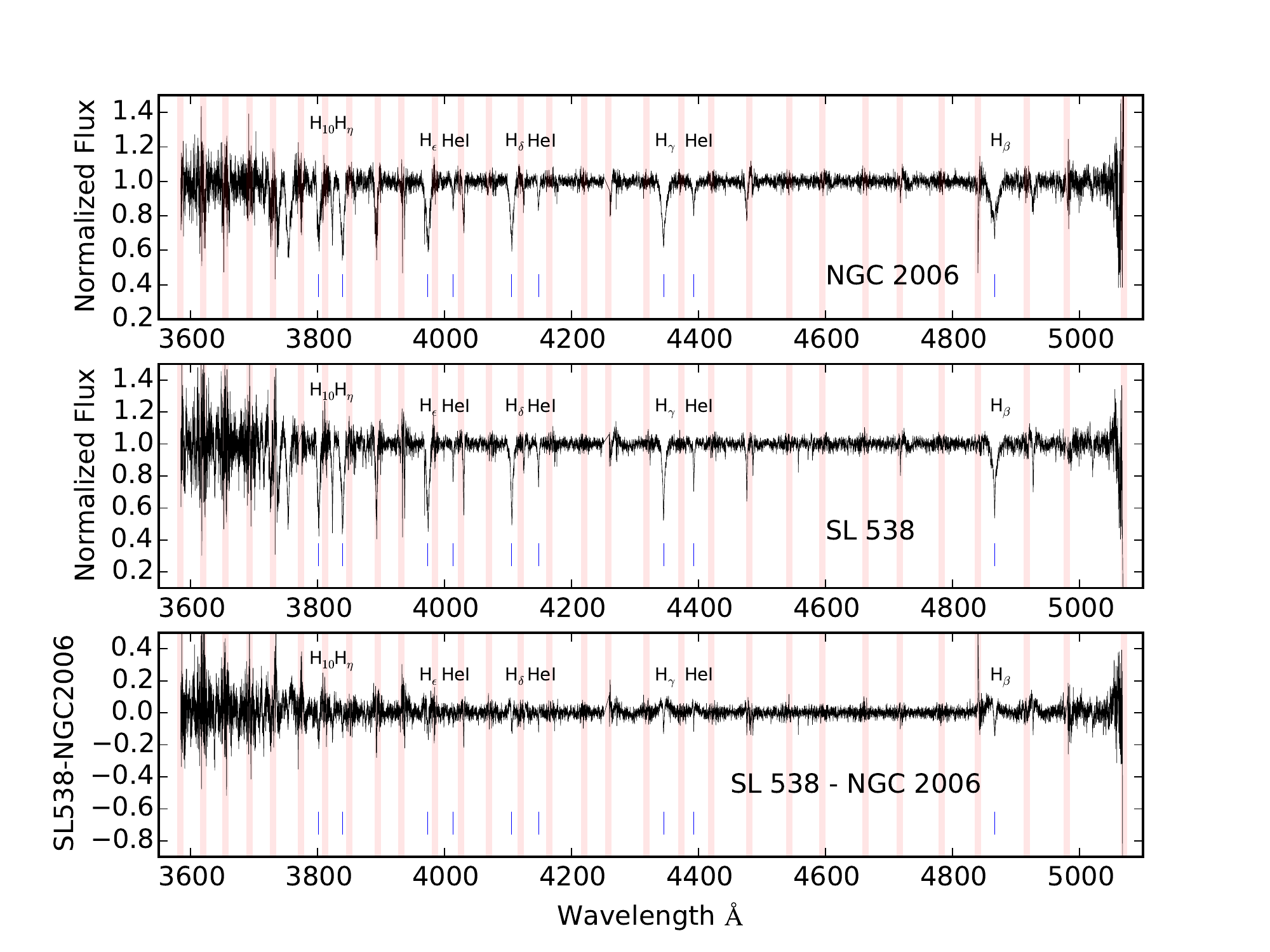} 
\caption{Normalized integrated-light spectra for NGC\,2006 (top panel) and SL538 (middle panel). The bottom panel shows the difference spectrum (SL538--NGC\,2006). The vertical red lines in each panel mark the echelle order overlap regions. Some strong absorption features are  labeled.}
\label{full_range}
\end{figure*}

The cluster spectra are consistent with a mixture of O-B type stars, in line with their approximate age of 20 Myr \citep{Dieball:1998rm, Kumar:2008rw}. We observe strong Balmer absorption and He\,{\sc i} absorption features (see Fig.~\ref{full_range}).~We, therefore, use the aforementioned features to accurately measure the cluster radial velocities and to assess the validity of the bound binary cluster scenario. The selection of the weaker absorption features is somewhat arbitrary and is based on their position (i.e.~at the central wavelength) within the corresponding echelle order. Since the transmission curves maximizes the S/N in approximately the center of each order along the dispersion axis (see Fig.~\ref{Curved}), features near the borders of the individual orders are rejected from the following analysis due to their small S/N.

\subsection{Evaluation of the line profile shapes}
In an attempt to select the best way of measuring the radial velocity from individual absorption features, we tested three different profiles: Gaussian, Moffat and Voigt. A first guess of the central wavelength position is measured using the {\sc Iraf} task \emph{splot}, which is then used as a starting point for the line profile fitting routine. The continuum near the absorption feature is mapped by a polynomial function of order 9, which is then subtracted from the spectra. Then the three profile types are fitted to the absorption features and the resulting fitting curve is subtracted from the data creating the residual spectra. This procedure is iterated until the solution converges (typically $3\times$). Finally, we choose the best fitting profile by visual inspection of the residuals and conclude that the residuals from the Voigt profile are smaller than the residuals from the Moffat and Gaussian profiles. Therefore, we adopt the Voigt profile for deriving central wavelength positions for our absorption features.

\subsection{Radial velocity measurements}
\subsubsection{Line profile fits}
\label{txt:lines}

\begin{table*}[t]
\centering
\caption[]{Summary of the radial velocity measurements from individual absorption features in NGC\,2006 and SL538.}
\begin{tabular}{lcccc}
\hline\hline
Cluster & Feature $\lambda_0$  &   $\lambda_m$  & $v_r$  & ${\Delta}v_{rms}$ \\
& \AA & \AA & $\mathrm{km}/\mathrm{s}$& $\mathrm{km}/\mathrm{s}$ \\
\hline\noalign{\smallskip}

NGC\,2006   &H$_{10}(3797.907)$   		& 3801.54$^{+0.10}_{-0.10}$ 	& 286.4$^{+8.4}_{-8.4}$ &$\pm$0.8\\
SL538	    &H$_{10}(3797.907)$   		& 3801.74$^{+0.05}_{-0.06}$ 	& 302.3$^{+4.1}_{-4.9}$ &$\pm$0.8\smallskip\\

NGC\,2006   &H$_{\eta}(3835.384)$  		& 3839.30$^{+0.12}_{-0.10}$ 	& 305.9$^{+8.0}_{-9.5}$ & $\pm$1.6\\
SL538	    &H$_{\eta}(3835.384)$  		& 3839.59$^{+0.07}_{-0.08}$ 	& 328.5$^{+6.3}_{-5.5}$&$\pm$1.6\smallskip\\

NGC\,2006   &H$_{\epsilon}(3970.072)$  		& 3973.73$^{+0.1}_{-0.1}$  	& 276.5$^{+7.7}_{-7.7}$&$\pm$0.8\\
SL538	    &H$_{\epsilon}(3970.072)$  		& 3973.91$^{+0.6}_{-0.5}$ 	& 289.5$^{+4.0}_{-4.8}$&$\pm$0.8\smallskip\\

NGC\,2006   &He\,{\sc i}(4009.256)  		& 4013.53$^{+0.06}_{-0.06}$ 	& 319.3$^{+4.8}_{-4.8}$&$\pm$3.3\\
SL538	    &He\,{\sc i}(4009.256)		    & 4013.53$^{+0.03}_{-0.04}$ 	& 319.6$^{+2.6}_{-2.9}$&$\pm$3.3\smallskip\\

NGC\,2006   &H$_{\delta}(4101.74)$  		& 4105.64$^{+0.09}_{-0.09}$ 	& 285.3$^{+6.5}_{-7.2}$&$\pm$1.5\\                         
SL538	    &H$_{\delta}(4101.74)$  		& 4105.96$^{+0.05}_{-0.03}$ 	& 309.1$^{+1.6}_{-3.8}$&$\pm$1.5\smallskip\\

NGC\,2006   &He\,{\sc i}(4143.761) 		& 4148.03$^{+0.09}_{-0.06}$ 	& 308.9$^{+4.2}_{-4.9}$&$\pm$1.5\\                        
SL538	    &He\,{\sc i}(4143.761)		& 4148.04$^{+0.02}_{-0.02}$ 	& 309.4$^{+1.6}_{-3.8}$&$\pm$1.5\smallskip\\

NGC\,2006   &H$_{\gamma}(4340.462)$  		& 4345.02$^{+0.05}_{-0.05}$ 	& 314.6$^{+2.9}_{-3.4}$& $\pm$2.8\\                           
SL538	    &H$_{\gamma}(4340.462)$  		& 4345.06$^{+0.02}_{-0.03}$ 	& 317.6$^{+2.2}_{-1.7}$& $\pm$2.8\smallskip\\

NGC\,2006   &He\,{\sc i}(4387.929)  		& 4392.46$^{+0.04}_{-0.04}$ 	& 309.4$^{+3.0}_{-3}$& $\pm$0.7\\                      
SL538	    &He\,{\sc i}(4387.929)  		& 4392.55$^{+0.02}_{-0.02}$ 	& 315.7$^{+1.2}_{-1.2}$& $\pm$0.7\smallskip\\

NGC\,2006   &H$_{\beta}(4861.342)$  		& 4866.39$^{+0.04}_{-0.04}$ 	& 311.2$^{+2.2}_{-2.2}$& $\pm$2\\  
SL538	    &H${_\beta}(4861.342)$  		& 4866.44$^{+0.01}_{-0.01}$  	& 314.4$^{+0.07}_{-0.07}$& $\pm$2\smallskip

\\
\hline\noalign{\smallskip}
NGC\,2006  & 						& \multicolumn{2}{r}{$\langle v_r\rangle=300.3\pm5 \pm5 ({\rm btstrp}) \pm6 ({\rm sys})$}\\
SL538          & 						& \multicolumn{2}{r}{$\langle v_r\rangle=310.2\pm4 \pm4 ({\rm btstrp}) \pm6 ({\rm sys}) $}\\
\hline\hline
\end{tabular}
\tablefoot{The provided radial velocity measurements are corrected to the heliocentric velocity ($\Delta v_h=-1.6$ km s$^{-1}$). The correction was calculated with the {\sc iraf} task \emph{rvcorrect} with the observatory set to LCO, for the observing time and coordinates of both targets. {{Please note that the radial velocity measurement differences seen on different lines are due to the wavelenght solution is not linear and thus values differ significantly more than the
uncertainties. Therefore focusing on the measured velocity difference between the clusters leaves
generally the relative values untouched while individual velocity measurements carry this systematic
uncertainty in their values. Moreover, due to the data reduction, each row of the straighten slit
was independently wavelength calibrate, therefore the differences are amplified during the spectra addition,
this systematic error is represented in the 5th  column and corresponds to the final rms. }}}
\label{velocidades_lines}
\end{table*}

For the measured central wavelength positions ($\lambda_m$) of the studied absorption features we derive the corresponding radial velocity values ($v_r$) and summarize the results in Table~\ref{velocidades_lines}. The corresponding line fits for both clusters are shown in Figure~\ref{voigt_fits} in the Appendix.~We observe a systematically larger radial velocity for SL538 compared to NGC\,2006 with an observed maximum $\lambda_m$ difference for the H${\delta}$ line of 0.32\,\AA, corresponding to a radial velocity offset of $\Delta v_r\!\simeq\!24$ km s$^{-1}$. We note that the largest offsets are measured for the lowest-S/N features with the largest uncertainties. Nevertheless, we observe that the large differences are consistent in the direction of the offset with the other features. At the bottom of Table~\ref{velocidades_lines}, we show the final mean radial velocities for each cluster based on the individual Voigt profile fits. 

In the following we estimate the reliability of our measurements by bootstrapping the science spectra with their variance properties and repeating the measurements. To assess the external errors of our measurements, we vary each pixel value according to its random noise component. For each pixel we randomize its original value ($N_{\rm sci}$) by adding a random component, assuming  $\sigma=\sqrt{g\cdot{\rm ADU}}$, where $g=0.47$ is the MIKE blue arm gain in $e^-$/ADU. Given the Poisson statistics and the original science spectrum a new artificial spectrum is generated, where for each pixel value we obtain $N_{\rm rnd} = N_{\rm sci} \pm P(\sqrt{N_{\rm sci}})$.~Here $P(\sqrt{N_{\rm sci}})$ is the Poisson distribution of each pixel value. We repeat this randomization and measurement procedure 10000 times. Visual examples of the fittings are presented in Figure~\ref{voigt_fits} and the measured radial velocity uncertainties are given in Table~\ref{velocidades_lines}, where the first error corresponds to the measurement uncertainties taking into account the variations of each line due to profile fitting.~The second error corresponds to the 95\% minimum and maximum recovered values, taken from our bootstrapping experiments, assuming a Gaussian distribution. This external error represents the reliability of the individual profile fitting, which is finally estimated as the resulting value.

We have found $v_r\!=\!300.3\pm5$ km s$^{-1}$ with a bootstrap error of 4 km s$^{-1}$ for NGC~2006 and $v_r=310.2\pm4$ km s$^{-1}$ and bootstrap error of 2 km s$^{-1}$ for SL538.~The first errors correspond to the standard error of the mean for the statistical uncertainty, while the second bootstrap error is discussed above.~Given our measurements we obtain a radial velocity difference estimate of $\Delta v_r\!=\!10\pm9\pm3$ km s$^{-1}$ ($\pm$stat$\pm$btstrp), which is away from no radial velocity offset but still consistent with it within 1 sigma uncertainties. 
~Note that the radial velocities for both clusters are corrected for the heliocentric radial velocity at the time of the observations of $\Delta v_h\!=\!-1.6$ km s$^{-1}$.~In general, SL538 shows systematically smaller errors than NGC\,2006 and its measurements are more stable for features at redder wavelength, which is a consequence of the MIKE transmission curve.

\subsection{Full spectrum cross correlation}
\label{txt:fullspec}
An alternative to the measurement of the central wavelengths of individual absorption lines is the full-spectrum cross correlation with radial velocity template spectra. Before applying the cross-correlation technique to the NGC\,2006 and SL538 spectra, we normalize each order and combine them into a master spectrum for each cluster showing the full wavelength coverage. Both spectra are shown in Figure~\ref{full_range} together with their corresponding difference spectrum. It is obvious that the blue and red edges of these spectra are relatively noisy compared to the central portion around $\sim\!3800\!-\!4900\,\AA$ which contain strong features well suited for cross-correlation.~We have used these portions of the spectra as input for a relative velocity estimation with the {\sc Iraf} task \emph{fxcorr}, which is based on the cross-correlation algorithm of \cite{Tonry:1979qy}.~Direct cross-correlation of both spectra, i.e.~using one as input science and the other as template spectrum, yields a relative radial velocity of $\Delta v_r=0.4$ km s$^{-1}$. 

\begin{figure}[b!]
\centering
\includegraphics[width=0.95\columnwidth,angle=0]{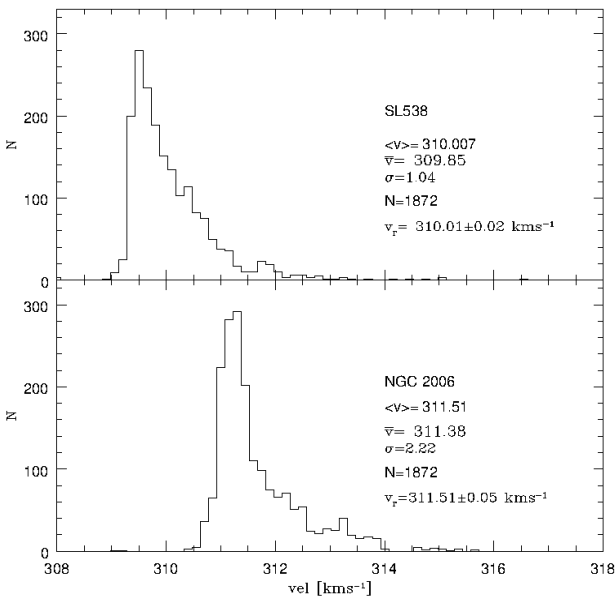} 
\caption{Distribution of derived radial velocities from cross-correlation of the two target clusters and 1872 synthetic stellar template spectra from \cite{Coelho:2014qy}.~Note the differences in the asymmetric distributions of the resulting radial velocities, which indicates the degree of template mismatch for each cluster-template spectrum combination.}
\label{vel_cohelo}
\end{figure}

Visual inspection of the normalized cluster spectra reveals some patterns that indicate that template mismatch might be a concern (see Fig.~\ref{full_range}). We, therefore, choose a stellar spectrum library as reference in order to evaluate such systematics. For this, we are using 1872 high-resolution theoretical stellar spectra from the library of \cite{Coelho:2014qy} as radial-velocity templates for the \emph{fxcorr} task.~Figure~\ref{vel_cohelo} shows the histogram of the derived radial velocity distributions from the cross-correlation of the cluster spectra with the template library. The velocity distribution histograms for both clusters show similarly asymmetric, but not identical shapes, which corroborates our suspicion that template mismatch is significant when auto-correlating the cluster spectra. We obtain for the two clusters an average radial velocity of $\langle v_r\rangle\!=\!311.51\pm 0.05$ km s$^{-1}$ and $310.01\pm0.02$ km s$^{-1}$ for NGC\,2006 and SL538, respectively, with a dispersion of $\sigma\!=\!1.04$ km s$^{-1}$ for SL538 and $\sigma\!=\!2.22$ km s$^{-1}$ for NGC\,2006, and corresponding median values of $\bar{v}\!=\!311.4$\,km s$^{-1}$ and $\bar{v}\!=\!309.9$\,km s$^{-1}$.~Considering the derived values from the cross correlation with the stellar template library we find a radial velocity difference between the two clusters of $\Delta v_r\!=\!1.50\pm0.07$ km s$^{-1}$. 

However, the asymmetry of the distributions in Figure~\ref{vel_cohelo} generated
by the asymmetric absorption profile shapes of some features, in particular H-epsilon which blends
with the Ca K+H lines. If template and object spectrum show different relative feature strengths this will
produce asymmetric offsets in radial velocity.  Moreover, this clearly shows that a proper selection of the standard star is critical, as a randomly chosen template may introduce velocity variations up to $\sim\!3\!-\!4$ km s$^{-1}$.~Histograms in Figure~\ref{vel_cohelo} also show the preferred values at their peaks, i.e.~the most frequent velocity value, given the spectral types included in the stellar spectra library \citep{Coelho:2014qy}. However, since the library has different spectral types frequencies, the peaks could just corresponds to the most frequent ones, so we performed 10000 random experiments where we randomly selected 5 templates for each stellar temperature and we look at the most common value, finding agreement between the peaks.~The standard star types at or near the peaks of the histograms in Figure~\ref{vel_cohelo} represent the best approximations to our cluster spectra.

\begin{figure}[t!]  
\centering
\includegraphics[width=1.\columnwidth,angle=0]{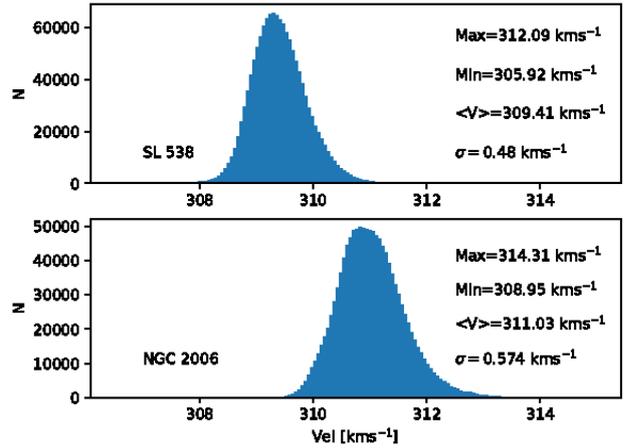}
\caption{Estimated radial velocity distributions for SL\,538 (top) and NGC\,2006 (bottom) from cross-correlation of the bootstrapped cluster and the best-template stellar library spectra (see Sect.~\ref{txt:fullspec} for details).}
\label{cluster_boot_synth}
\end{figure}

To assess the influence of the S/N of our cluster spectra independent of the template mismatch error, we have selected the best 9 stellar library  spectra, corresponding to stars of temperatures 5750-6000K and  consistent to be present in the peaks of  our random experiments as well as in the Fig. Figure~\ref{vel_cohelo}, as template radial velocity standards.~Each template was transformed to the resolution of the observed spectra, and had proportional noise added, based on the S/N of each observed cluster spectrum.~Finally the pixel values of template and cluster spectra were bootstrapped 150000 times according to the variance spectra of the corresponding cluster and the relative radial velocity was measured using cross-correlation.~Values were recorded and merged into a single histogram which is presented in Figure~\ref{cluster_boot_synth} for each cluster.~The histograms resemble a Gaussian distribution, but we observe a positive skewness for both clusters (i.e.~a slightly more extended right tail) and a slightly platykurtic distribution for NGC\,2006 (bottom panel).~Therefore, template mismatch may still be a systematic in this exercise, but is affecting our measurements significantly less than using the whole stellar library.~In conclusion, we adopt the following values from these experiments: For NGC\,2006 we obtain $v_{r}\!=\!311.0\!\pm\!0.6$ km s$^{-1}$ and for SL\,538 we measure $v_{r}\!=\!309.4\!\pm\!0.5$ km s$^{-1}$ where the uncertainties correspond to 1$\sigma$ of the distributions.~The minimum and maximum values of the distributions are given in the corresponding panel.~Hence, the S/N of our spectra limits our radial velocity measurements to a statistical accuracy of $0.5-0.6$ km s$^{-1}$.

\begin{figure}[h!]
\includegraphics[width=\columnwidth,angle=0]{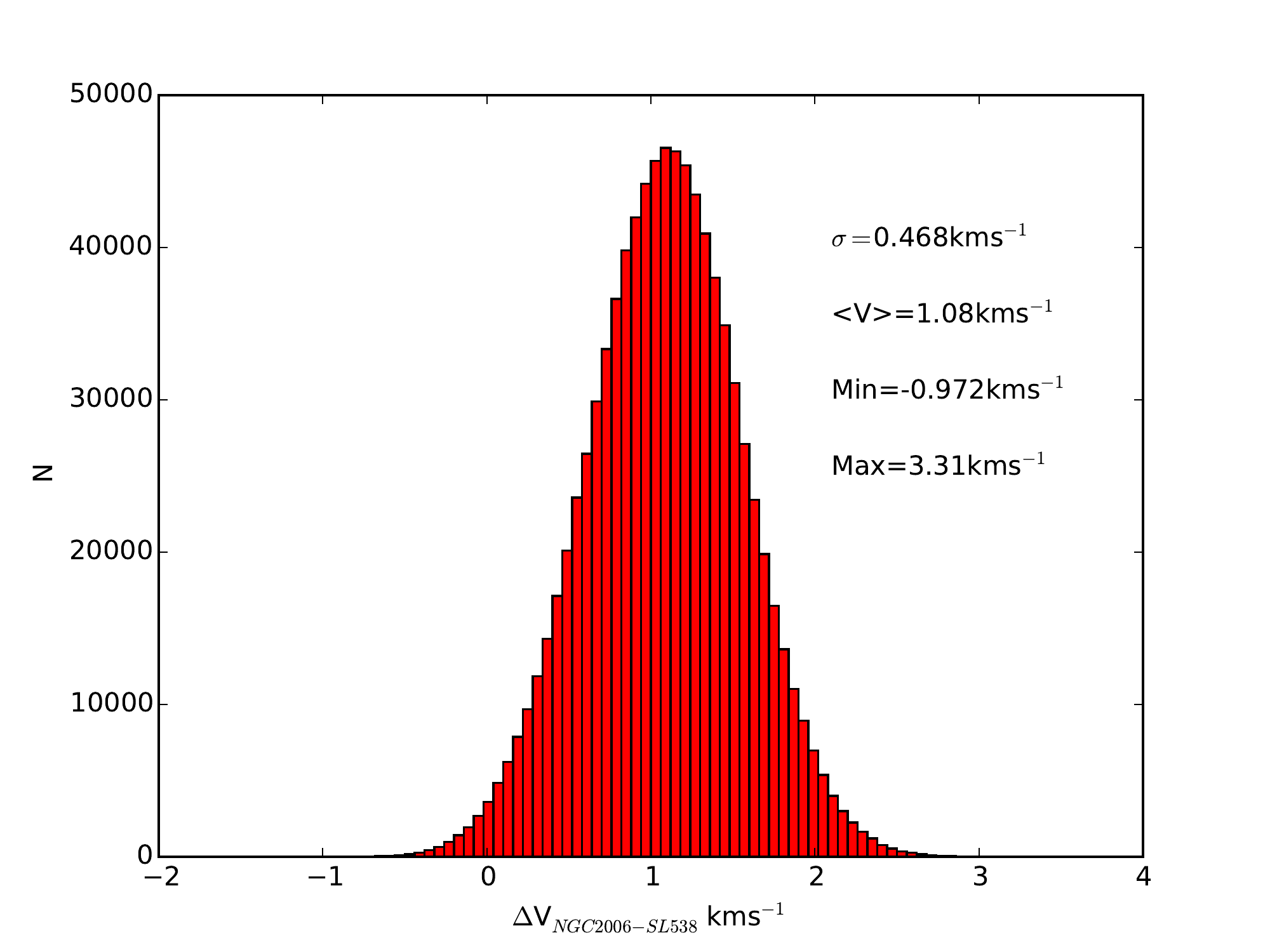} 
\caption{Distribution of the radial velocity difference between SL\,538 and NGC\,2006 obtained by direct cross-correlation of their bootstrapped spectra (see Sect.~\ref{txt:fullspec}) for details).}
\label{cluster_cross}
\end{figure}

To finalize our radial velocity uncertainty assessments, we now estimate the bootstrapped error for the earlier implemented direct cross-correlation of both cluster spectra. We following the same bootstrapping procedure described earlier to alter the spectra for each step according the the S/N of the spectra.~The resulting distribution is presented in Figure~\ref{cluster_cross}, from which we measure $\Delta v_r\!=\!1.08\pm0.47$ km s$^{-1}$ as the adopted radial velocity difference between the two target clusters.~The velocity difference distribution is consistent with our earlier experiments using the spectral library spectra to assess the spectral template mismatch and falls within the $\Delta v_r\!=\!10\pm9\pm6$ km s$^{-1}$ range obtained from the individual line measurements (see Sect.~\ref{txt:lines}).  

\subsection{Relative flux comparison}
Besides comparing the star-cluster radial velocities, we investigate, in the following, the differences in the stellar population contents based on selected, strong spectral absorption features. We compare both integrated spectra by subtracting them from one another and analyzing the residuals (see Fig.~\ref{full_range}).~Despite the fact that during the correction and combination of the individual orders we activated the flux conservation option in the {\sc Iraf} package, we pay particular attention to the echelle order overlap regions, which are marked in Figure~\ref{full_range} by vertical red stripes.~Since the data reduction process was done identically for the two spectra, each one of them is affected by the same systematics.~Therefore, by dividing/subtracting one by/from the other, the systematics will not affect our comparison as each spectrum follows the corresponding transmission curves of each order identically. Because of the obtained moderate data quality (peak S/N~$\!\simeq\!13.6$ for NGC\,2006 and $\sim\!10.7$ for SL\,538), we have decided not to fit stellar population model predictions to our spectra, but to compare qualitatively the shape of the absorption features from Table~\ref{velocidades_lines} for indications of different abundances.

The comparison was done in the following way: For each cluster we bootstrap the spectrum according to its variance spectrum, (i.e.~adding noise in the same way as it was done for the radial velocity uncertainty assessments; see previous sections).~Then, we normalize the bootstrapped orders and finally subtract and divide the normalized fluxes of each cluster for a direct comparison.~The results are presented in Figure~\ref{Comparison} and show that there are significant differences in the line strengths of helium and the Balmer series between NGC\,2006 and SL\,538.~In the case of hydrogen, the NGC\,2006 spectrum shows stronger Balmer absorption, while the residuals are very symmetric for each of those absorption features.~This is in line with the expectations from the CMD ages found for NGC\,2006 of $22.5\!-\!25$ Myr, with SL\,538 being $\sim\!5$ Myr younger.~However, the helium lines show indications of substructure in the difference and flux-ratio spectra (see Fig.~\ref{Comparison}), possibly hinting at different radial velocity distributions of stars (incl.~stellar rotation velocity distribution profiles, stellar rotation axis alignments, stellar mass segregation, etc.), which contribute mostly to these features with atmospheres typically hotter than those of B-type stars.~We require higher S/N spectra to investigate the details of this intriguing difference in the absorption profiles of the various element species between these two star clusters.

\subsection{Stellar population ages}
In an attempt to constraint ages and  to compare our observations with the predictions, and to roughly estimate metallicities with the aim to explore the differences (or similarities) that may hint the possible formation scenario and to compare our work with previous measurements, we have used high-resolution spectral energy distribution (SED) models from Starburst99 \citep[v7.0.1][]{Leitherer:1999qy,Leitherer:2010kx,Leitherer:2014ys,Vazquez:2005vn}.~Each normalized cluster SED was compared against theoretical SEDs (also normalized) with ages spanning from 5~Myr up to 40~Myr with intermediate steps of 1~Myr, considering the following metallicities: $Z\!=\!0.001, 0.008, 0.02$, and $0.04$ dex.~We used the spectral region between $3750$ \AA\ and $5000$ \AA, avoiding the orders with the highest noise.
 
\begin{figure}[t!]
\includegraphics[width=\columnwidth,angle=0]{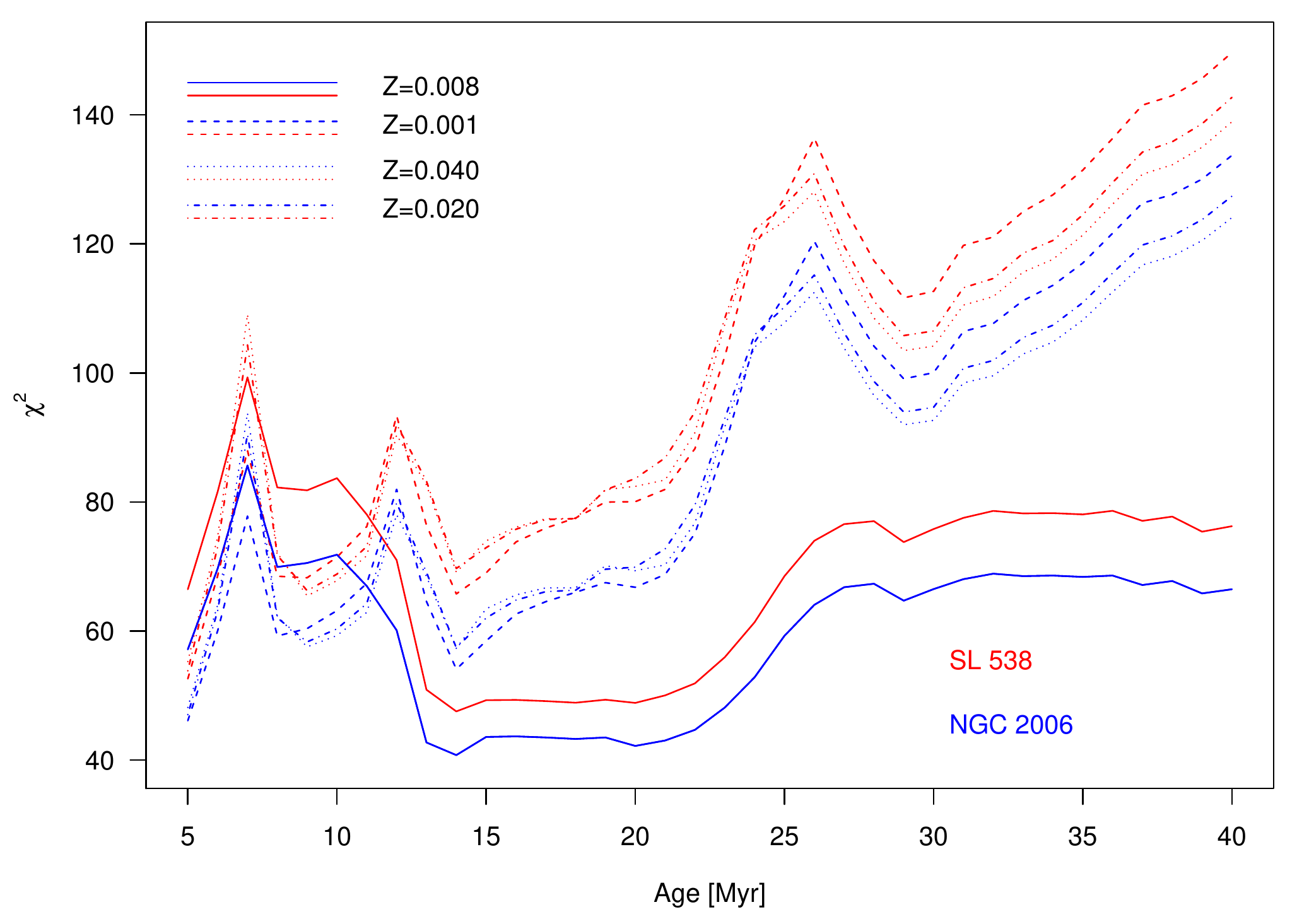} 
\caption{$\chi^2$ as a function of stellar population age for a comparison between {\sc Starburts99} SEDs and the integrated spectra of NGC\,2006 (blue) and SL\,538 (red).~Various lines types encode the SED metallicity with values for $Z\!=\!0.001, 0.008, 0.02$, and $0.04$ as explained in the top-left legend.~The absolute minimum is located for both clusters at 14 Myr for a metallicity of $Z\!=\!0.008$, however, from the shape of the $\chi^2$ curve, the minimum is likely in the region between 14 and 21 Myr.}
\label{chi2}
\end{figure}

\begin{figure}
\includegraphics[width=\columnwidth,angle=0]{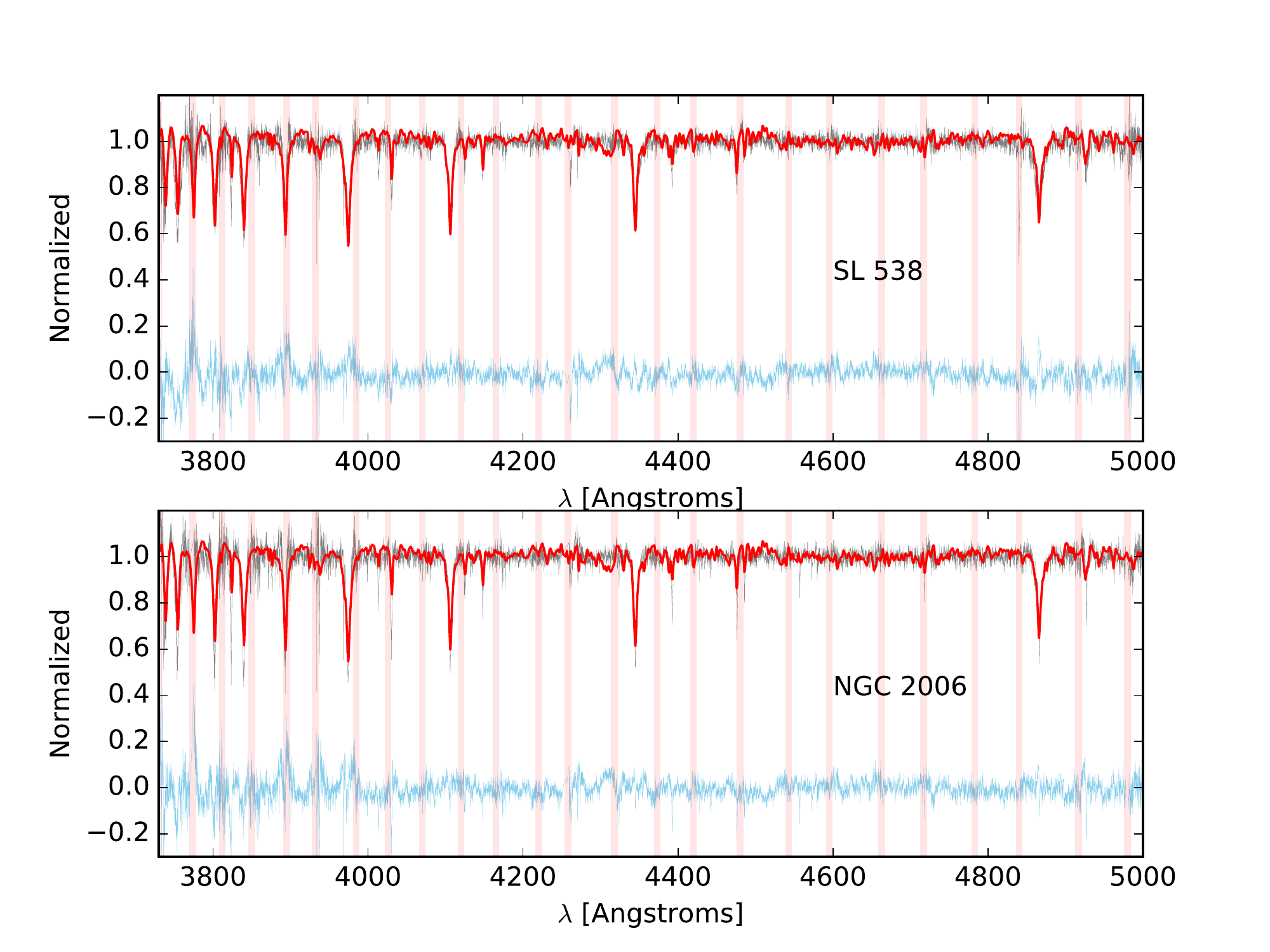} 
\caption{Best model fit corresponds to an age of 14 Myr for both clusters. On each panel, the normalized observed spectrum is shown in grey, the red line corresponds to the Starburst99 14~Myr, model with a metallicity of $Z=0.008$, and in light blue the residual (i.e. spectra subtracted by the model) is shown. Red stripes corresponds to the limits of the individual Echelle orders.}
\label{BEST_FIT}
\end{figure}

Results are presented in Figure~\ref{chi2} where we show the $\chi^{2}$ as function of age for various metallicities.~Residuals were visually inspected and we found that the absorption lines were not well reproduced by SEDs of 5 Myr.~For the case of $Z\!=\!0.008$ metallicity, absorption lines were best fitted (within the noise of the data) by SEDs with ages in the range $\sim\!13\!-\!21$~Myr with the formal absolute $\chi^{2}$-minimum at 14 Myr. The best-fit SED and residuals are shown in Figure~\ref{BEST_FIT}.~Due to the limited S/N of our data we were not able to further constrain the ages and metallicities of both clusters.~Despite the formal minimum $\chi^2$ favoring an age of 14 Myr, we consider that the stellar population ages of both clusters fall in a range between 13 and 21~Myr for a metallicity near $Z\!=\!0.008$, with a potential indication for the existence of very young, i.e.~5 Myr old stars.~We point out that this is consistent with the results from \citep{Dieball:1998rm}.


\section{Discussion}
\label{ln:disc}
\subsection{Cluster binarity}
Measuring radial velocities on individual absorption features and the full spectrum gives us an idea of the robustness and stability of our measurements.~Depending on the S/N of the spectrum at a given wavelength, we observe that the radial velocity values become more dispersed for blue features than for red ones.~We also note that the radial velocities measured with the two techniques are more consistent for SL\,538 than for NGC\,2006, which is likely due to the noise level of each individual profiles and their wider profile shapes.

For the following calculations we use the integrated-light measurements from the SIMBAD database\footnote{http://simbad.u-strasbg.fr/simbad/sim-fxxxx}, which give $V\!=\!10.88$\,mag for NGC\,2006 and $V\!=\!11.30$\,mag for SL\,538, and assume a distance modulus of $(m\!-\!M)\!=\!18.493$ \citep{Pietrzynski:2013rz} to both clusters and an absolute $V$-band solar luminosity of $M_{V\sun}\!=\!4.82$ mag.~We estimate a luminosity of $L_V\!=\!9.3\!\times\!10^4\,L_{\sun}$ for NGC\,2006 and $L_V\!=\!6.9\times10^4\,L_{\sun}$ for SL\,538.~If we adopt a mass-to-light ration of $M/L_{V}\!=\!0.064$ for a simple stellar population of 30 Myr (youngest age available in the model) and a metallicity $Z\!=\!0.008$ with solar-scaled chemical composition from the BaSTI database\footnote{http://albione.oa-teramo.inaf.it/}, this yields a total stellar mass of $M_{\rm cl}\!=\!5.9\times10^{3}\,M_{\sun}$ for NGC\,2006 and $4.3 \times10^{3}\,M_{\sun}$ for SL\,538.

If we assume a circular orbit and a statistical proxy of the de-projected three-dimensional distance $R_{\mathrm{3D}}\!=\!R_{P}\sqrt{3/2}$, where $R_P$ is the projected distance ($R_{P}\!=\!13.3$ pc), the orbital period in years is then: $P_{\rm orb}\!=\!9.3\times10^{7} R_{\mathrm{3D}}^{3/2}(m_1+m_2)^{-1/2}$, which yields $P_{\rm orb}\!\simeq\!60$\,Myr.~Therefore, the orbital velocity on a circular orbit can be computed as $v_{\rm orb}\!=\!2{\pi}R_{\mathrm{3D}}/P_{\rm orb}$, which results in $v_{\rm orb}\!=\!1.66$ km s$^{-1}$. A similar result is obtained if values from Table~1 from \cite{Fujimoto:1997uq} are taken into account. We point out that this value corresponds to the maximum observable velocity difference between the two clusters, but can change if elliptical orbits are invoked.~In any case, the computed value agrees with our observed radial velocity difference of a few km s$^{-1}$. Based on this estimate and similar, observed radial velocity differences in other cluster pairs, numerical simulations of the dynamical evolution of such star-cluster complexes \citep[e.g.][]{Yu:2017, Gudric:2017} suggest that these cluster pairs will merger within a few Myr.~Based on their numerical simulations, \cite{Bruens:2011} find that about half of the star cluster in their hierarchical complexes have merged after 2.5 crossing times, i.e.~$2.5t_{\rm cr}$.~For our binary system we obtain
\begin{equation*}
t_{\rm merge}\simeq2.5t_{\rm cr}=2.5\left(\frac{3\pi}{32}\right)^{-3/2}\sqrt{\frac{R_{\mathrm{3D}}^3}{GM_{\rm cl}}}\simeq150\, {\rm Myr}
\end{equation*}
using the values for $R_{\mathrm{3D}}$ and the total mass of the binary cluster ($10^4\,M_\odot$).~Together these results argue in favor of the scenario in which such clusters are gravitationally bound entities, i.e.~true binary star clusters that will merge within a few orbital timescales.~However, we recall that absorption features are not identical in both clusters (see Sect.~\ref{txt:fullspec}), which suggests a slightly different chemical abundance composition, at least in the case of helium.~This may hint at a non-negligible variance and/or non-isotropy during the clusters' (self-)enrichment process.

\subsection{Formation scenario}
Theoretical work from \cite{Fujimoto:1997uq} advocates that the formation of binary star clusters may be a result of oblique cloud-cloud collisions that give rise to cloud fragmentation (i.e.~Jeans gas clumps) and produce two or more gravitationally bound star clusters of roughly the same age in orbit with each other. This is, for example, believed to be the case in the binary cluster system NGC\,2136/2137 \citep{Mucciarelli:2012fk}.~A different scenario has been proposed by \citep{Leon:1999sf} where binary star clusters are born in rather separate environments, and form later by tidal capture, which would imply the presence of binary star clusters with significant age differences of their constituent stellar populations or a synchronized star-burst episode on spatial scales larger than individual proto-cluster complexes if no such age differences are measured.~The derived CMD ages for our target clusters NGC\,2006/SL\,538 measured by \citep{Kumar:2008rw}, \citep{Dieball:1998rm} and the similar ages derived from SED in our work show that overall an age difference between the two clusters is of the order of $\lesssim\!5$ Myr.~Together with derived dynamical properties and the direct comparison of absorption line profiles of different elements showing some interesting residuals, which can be interpreted as a difference in internal chemical composition of each cluster, we tend to favor a scenario in which the origin of the NGC\,2006/SL\,538 pair is situated in a loosely bound star-formation complex that later on becomes a binary star cluster through tidal capture \citep[see e.g.][]{Gudric:2017}.~Because of their similar masses and given the current separation and dynamics, this cluster pair is expected to merge within $\sim\!150$ Myr to form an open-cluster equivalent of approximately $10^4\,M_{\sun}$ that will dissolve over the next few Gyrs.

\section{Conclusions}
The analysis of the derived velocity differences of the star cluster pair NGC\,2006/SL538 presented in this study are consistent with the expected orbital velocity of a gravitationally bound binary star cluster pair that will merge with the next $\sim150$ Myr.~Cluster ages were constrained through full-SED fitting, but due to the low signal to noise, determined to lie in a range between $13$ and $21$ Myr for both clusters.~When the normalized, integrated-light spectra are compared, we see some differences in the residuals of absorption features from helium, which may be interpreted as differences in their chemical composition, concluding that this is likely a binary cluster that formed via tidal capture within a loosely bound star-formation complex rather than fragmentation of the same parent gas cloud.

\begin{acknowledgements}
We thank the anonymous referee for his/her comments that help us to improve this paper.
M. D. Mora is supported by CONICYT, Programa de astronom\'ia, Fondo GEMINI, posici\'on Postoctoral.THP acknowledges support through a FONDECYT Regular Project Grant (No. 1161817) and the support from CONICYT project Basal AFB-170002.
%
\end{acknowledgements}


\bibliographystyle{aa}
\bibliography{Marcelo_latest}

\begin{appendix}
\section{Absorption profile fits}

\begin{figure*}[!ht]
\centering
\includegraphics[width=0.49\textwidth,angle=0]{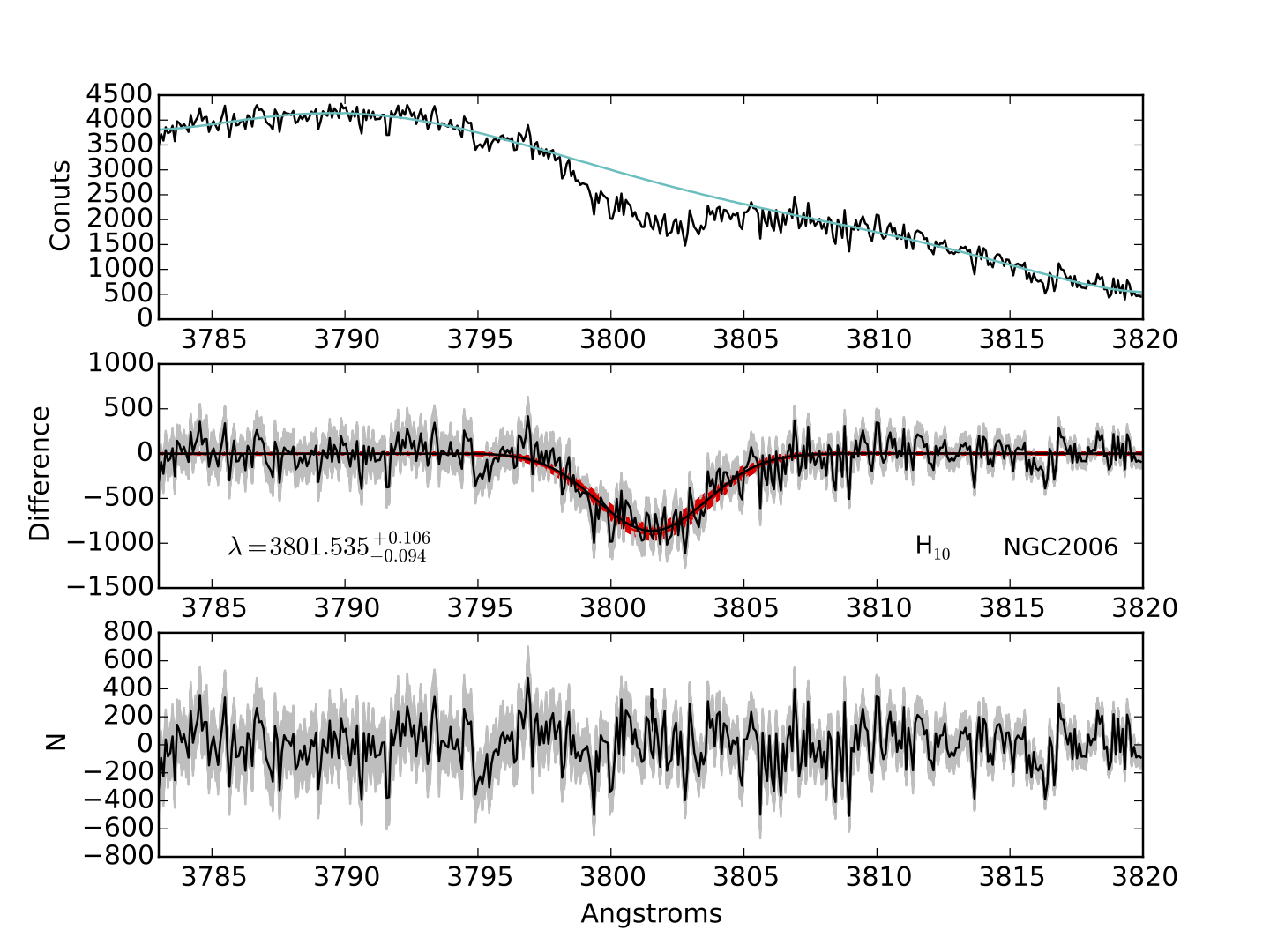}
\includegraphics[width=0.49\textwidth,angle=0]{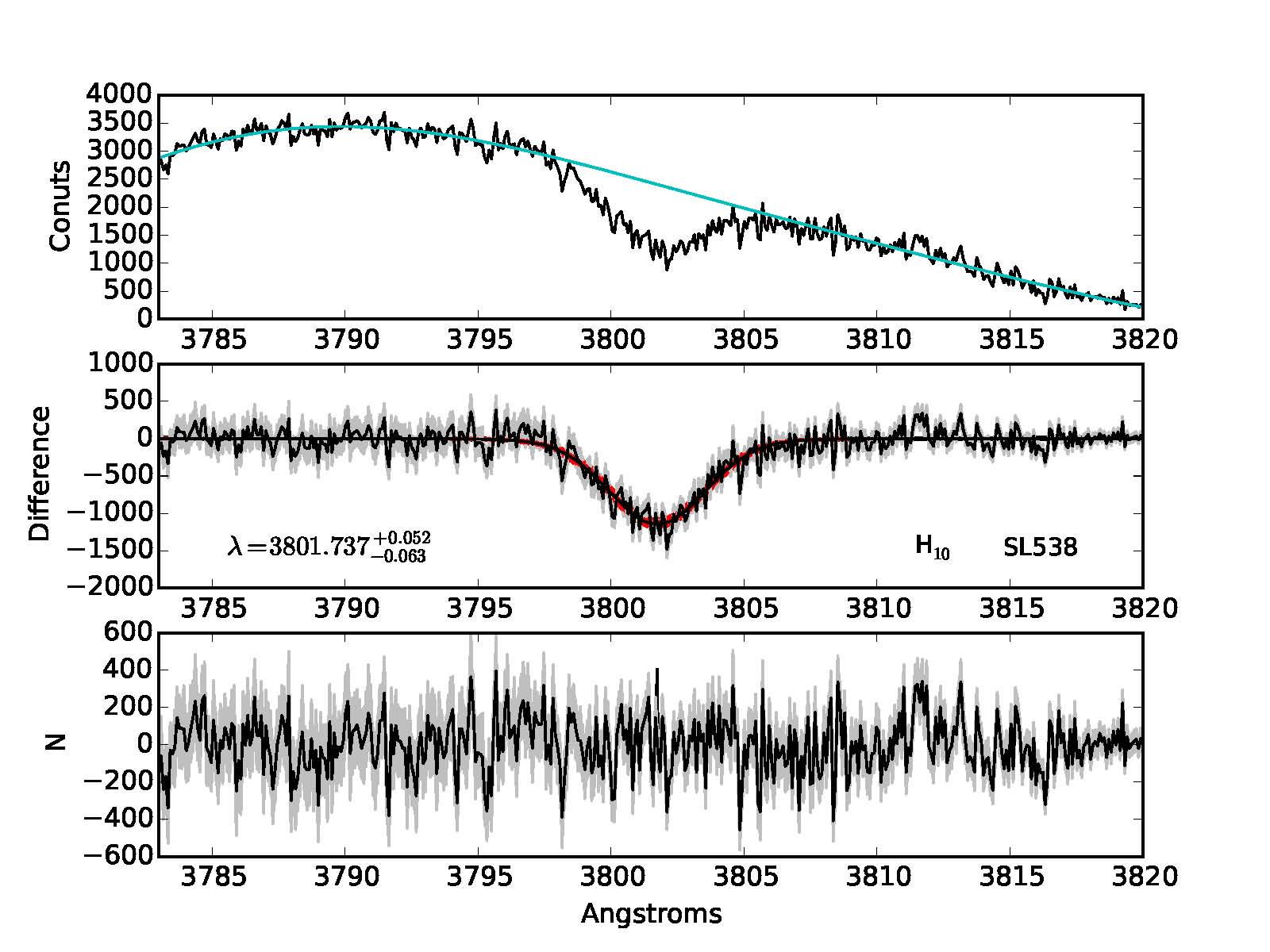}
\includegraphics[width=0.49\textwidth,angle=0]{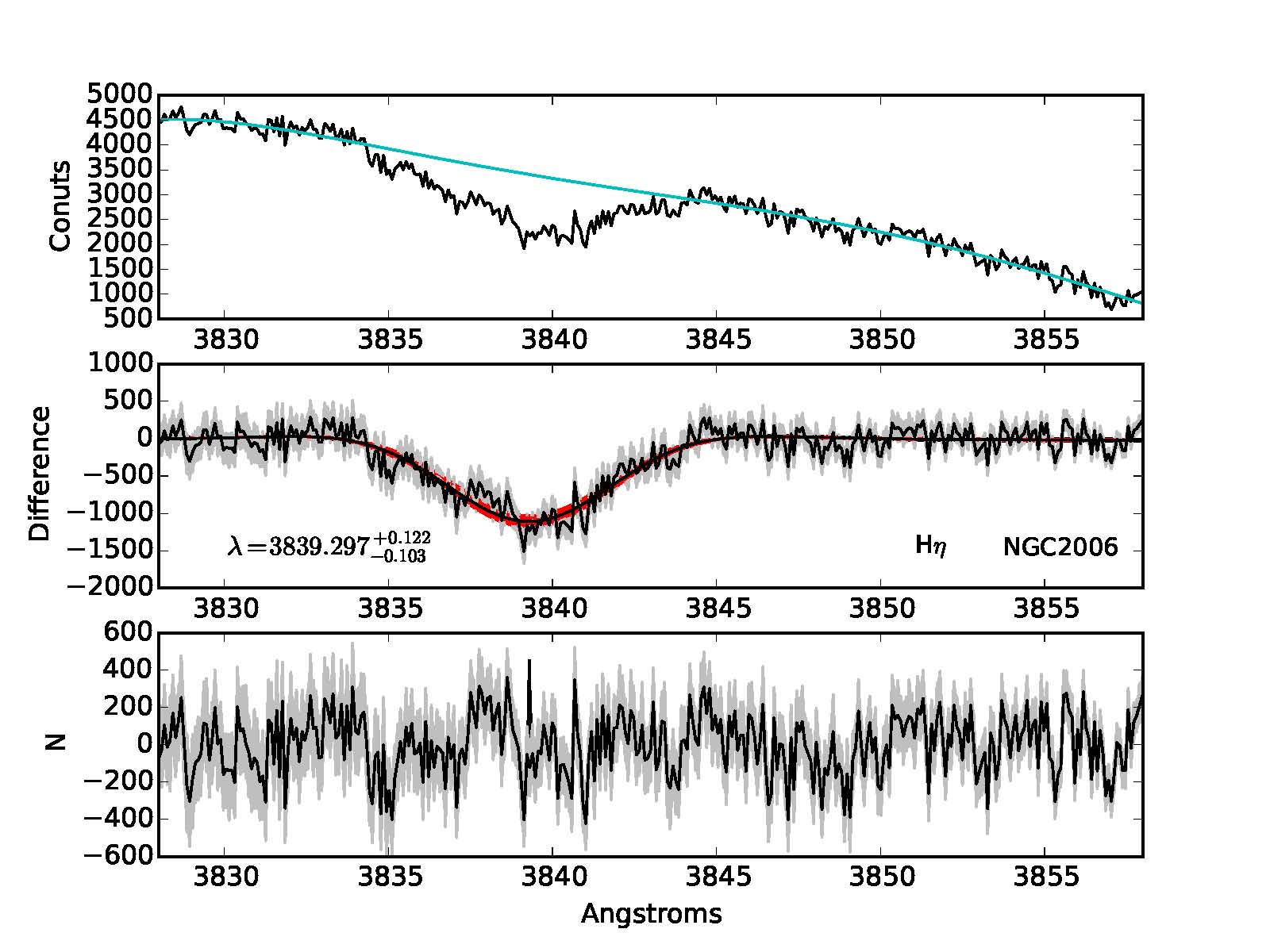}
\includegraphics[width=0.49\textwidth,angle=0]{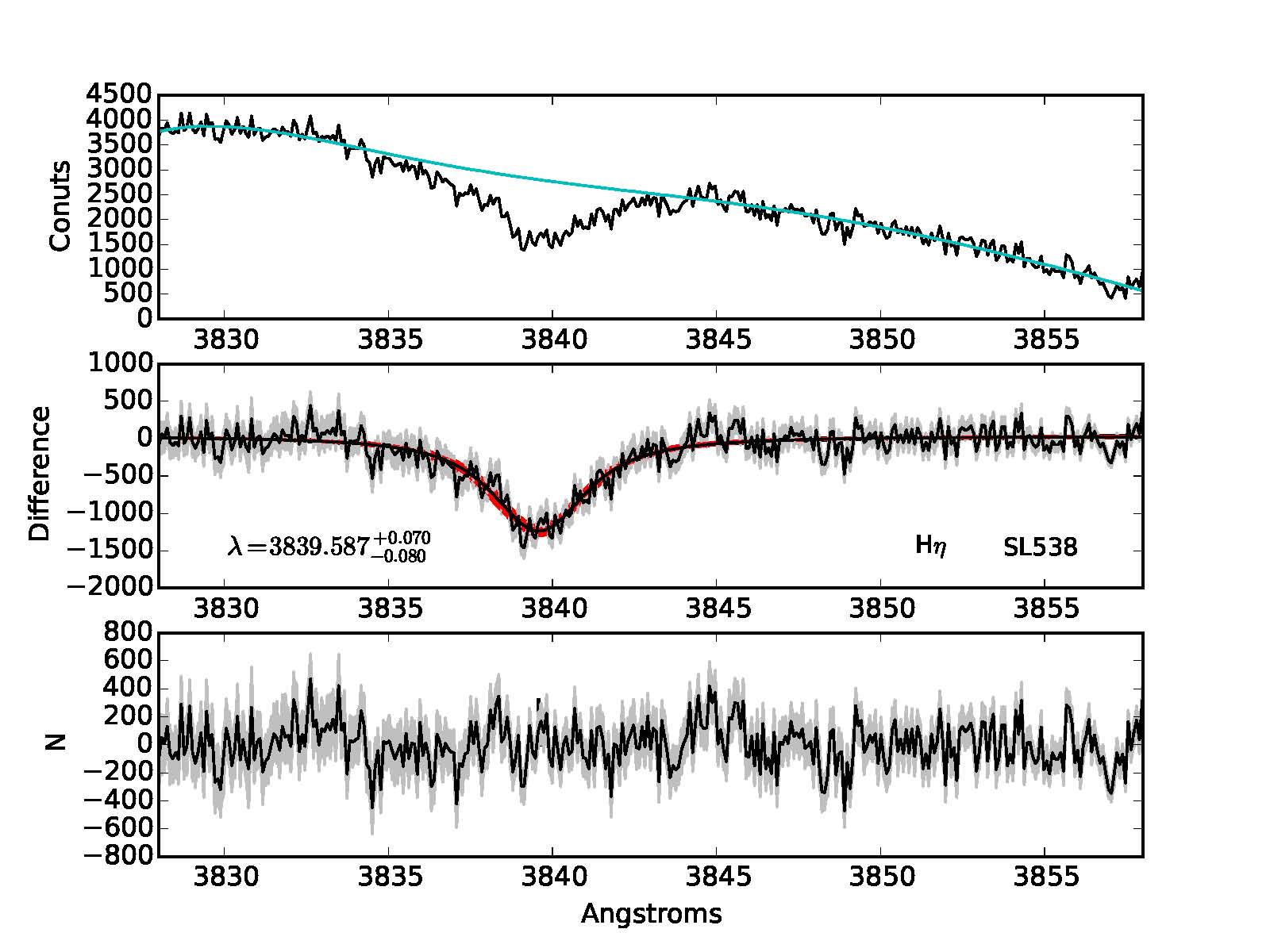}
\includegraphics[width=0.49\textwidth,angle=0]{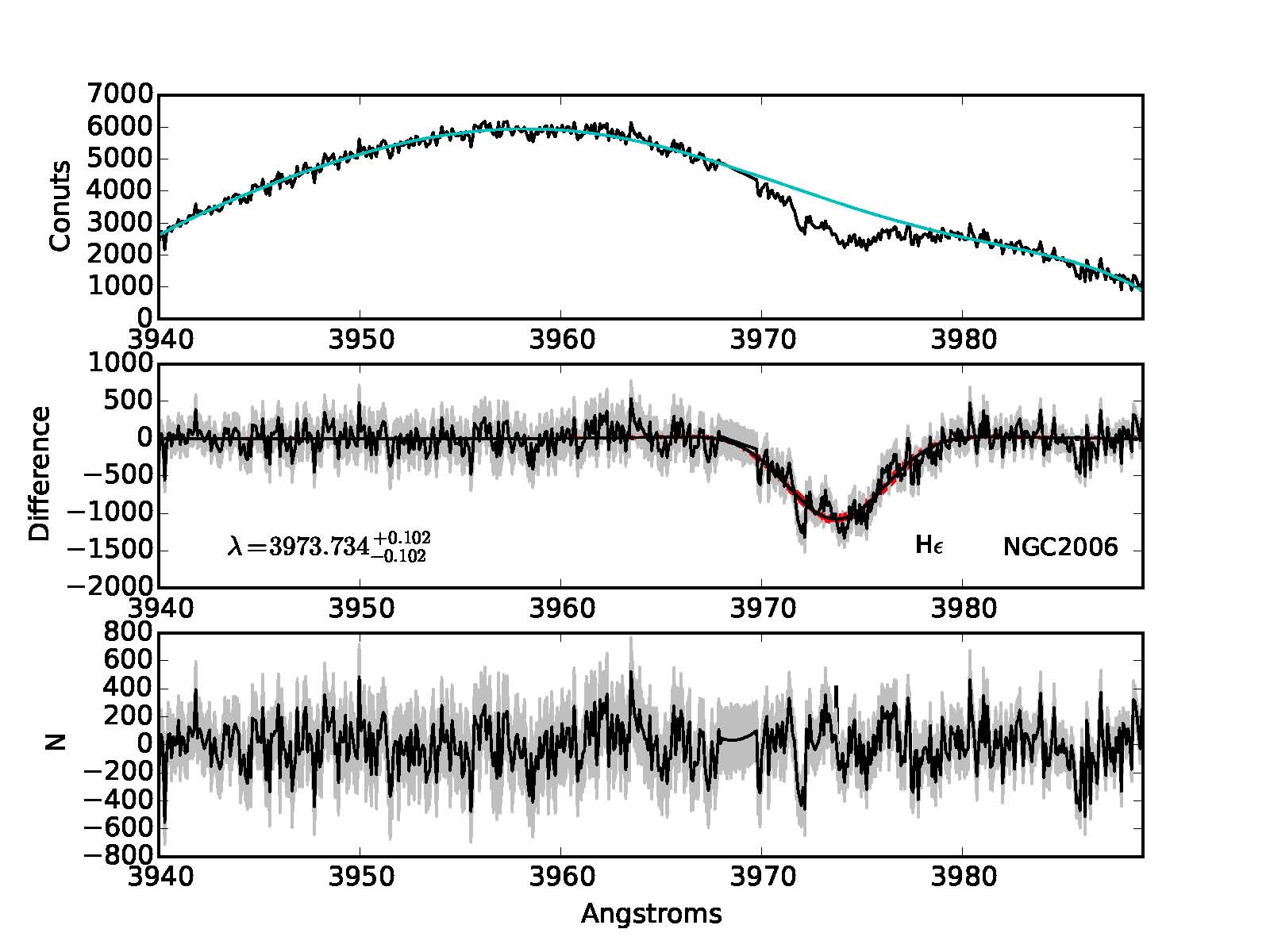}
\includegraphics[width=0.49\textwidth,angle=0]{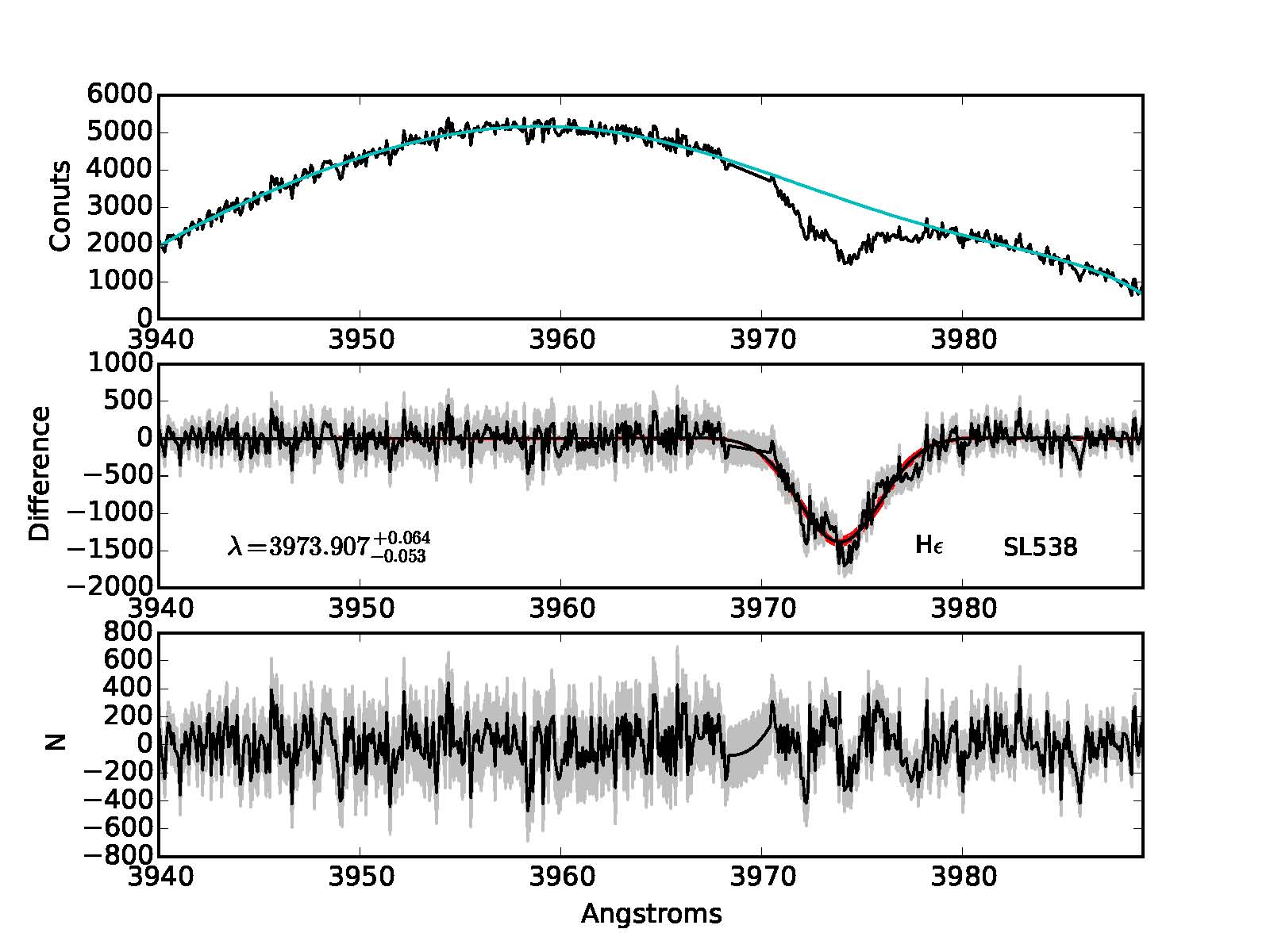}
\caption{Absorption line profile fits for all features listed in Table~\ref{velocidades_lines}. Features for the NGC\,2006 spectrum are shown in the left panels, while the corresponding fits for SL\,538 are given in the right panels. Each plot has three sub-panels: The upper one shows the observed spectrum near the absorption feature, where the cyan curve is the polynomial approximation to the continuum. The central sub-panel corresponds to the spectrum with the continuum subtracted, in which the black curve illustrates the Voigt profile fit, while the grey shadow represents the expected uncertainties from our bootstrapped experiments (see Section~\ref{txt:lines}). The red curves in the central sub-panel correspond to the bootstrapped 1$\sigma$ uncertainty of the profile fit. The lower sub-panels show the residuals with their after continuum and profile subtraction together with their 1$\sigma$ uncertainties as grey shadows. This figure shows profile fits for Balmer absorption features H$_{10}$, H$_{\eta}$ and H$_{\epsilon}$.}
\label{voigt_fits}
\end{figure*}

\addtocounter{figure}{-1}
\begin{figure*}
\centering
\includegraphics[width=0.49\textwidth,angle=0]{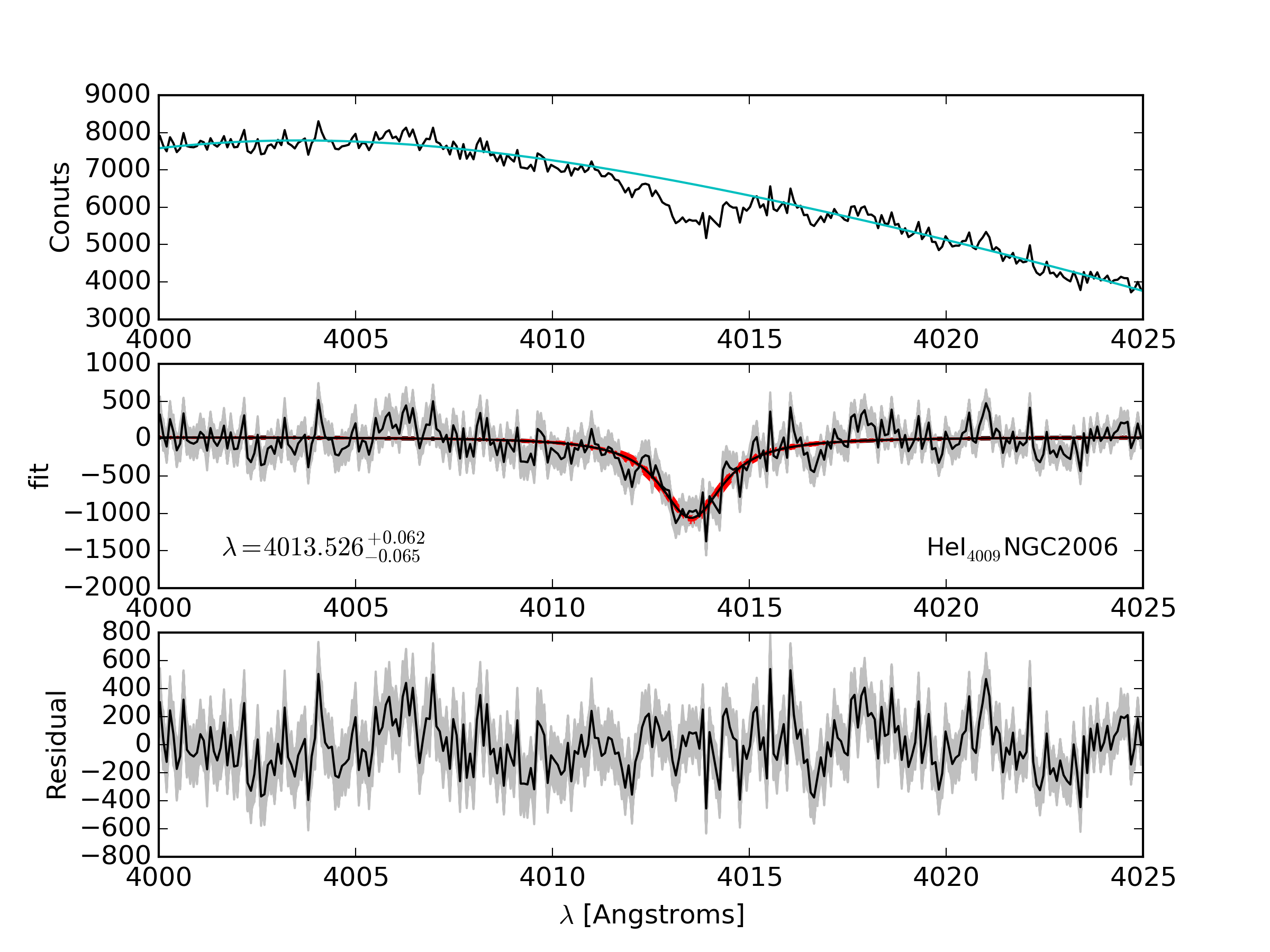}
\includegraphics[width=0.49\textwidth,angle=0]{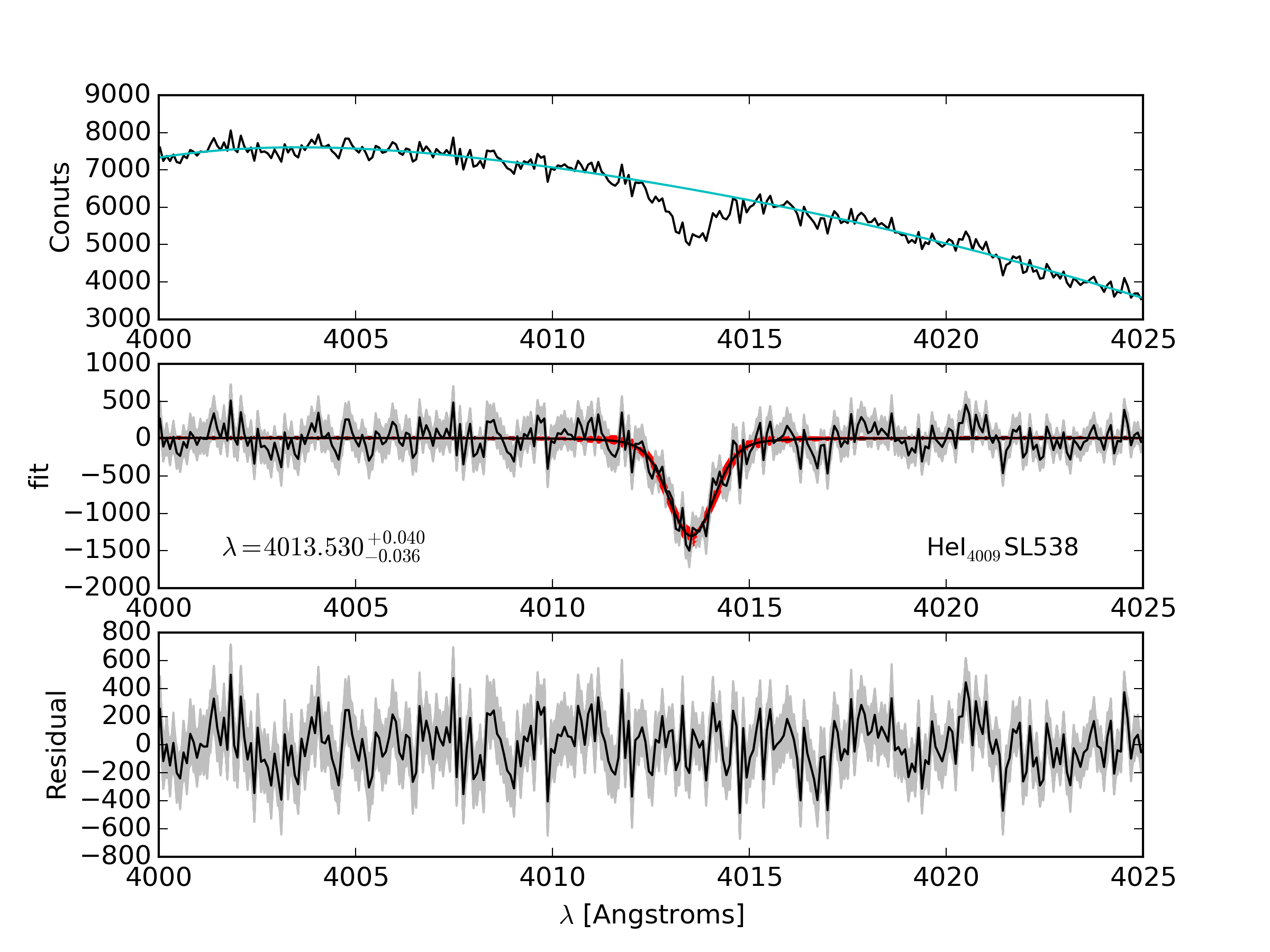}
\includegraphics[width=0.49\textwidth,angle=0]{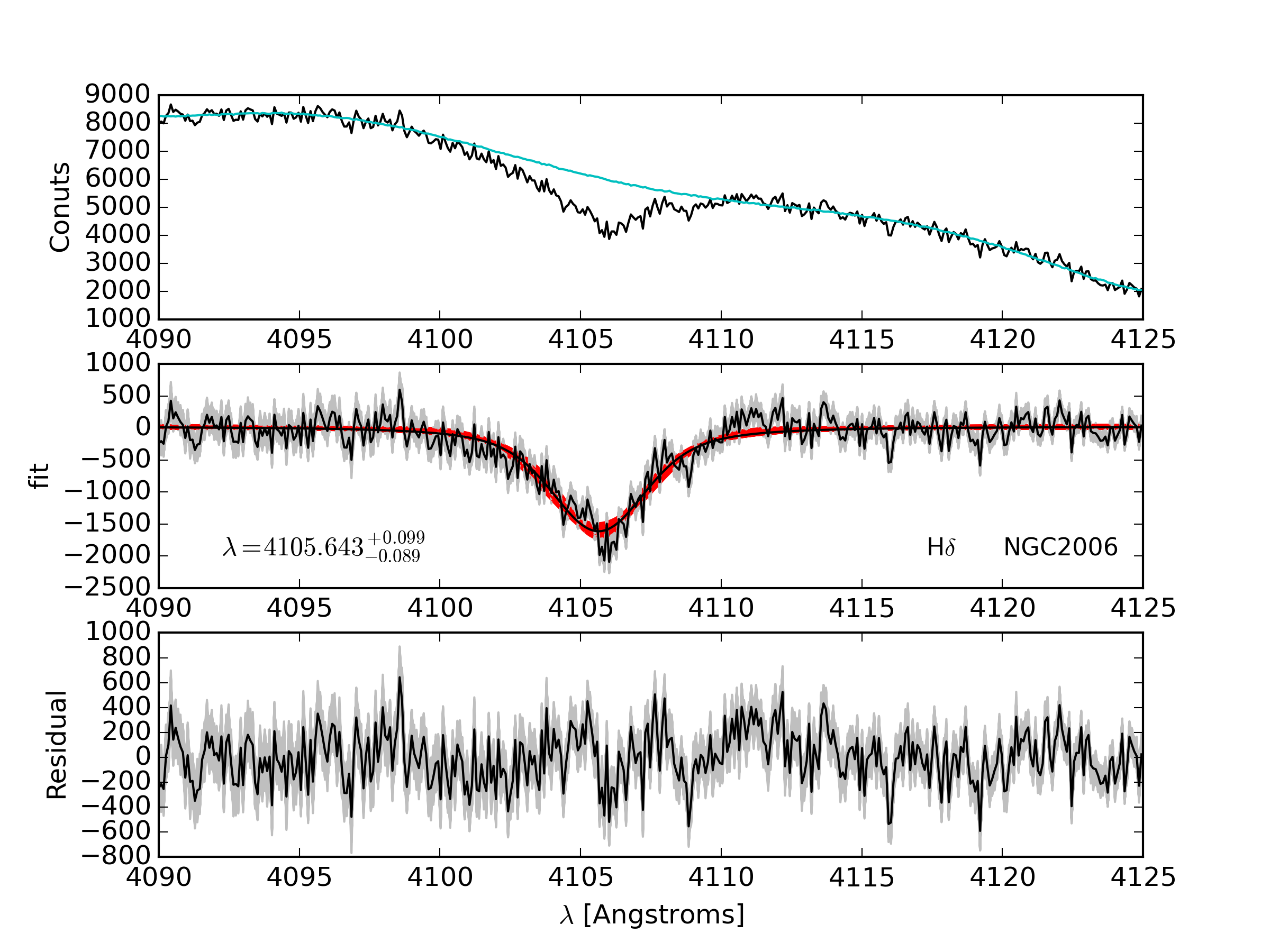} 
\includegraphics[width=0.49\textwidth,angle=0]{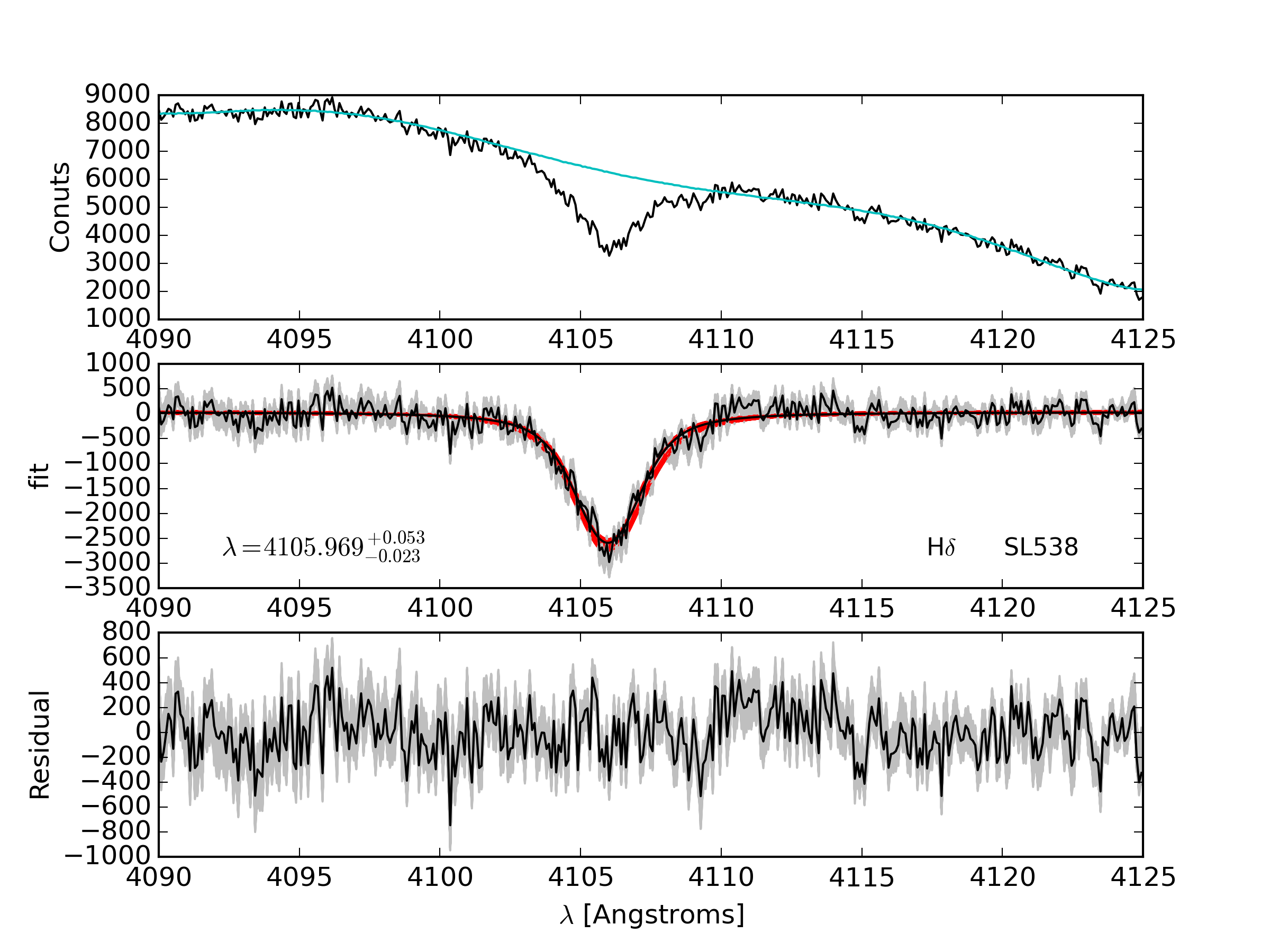}
\includegraphics[width=0.49\textwidth,angle=0]{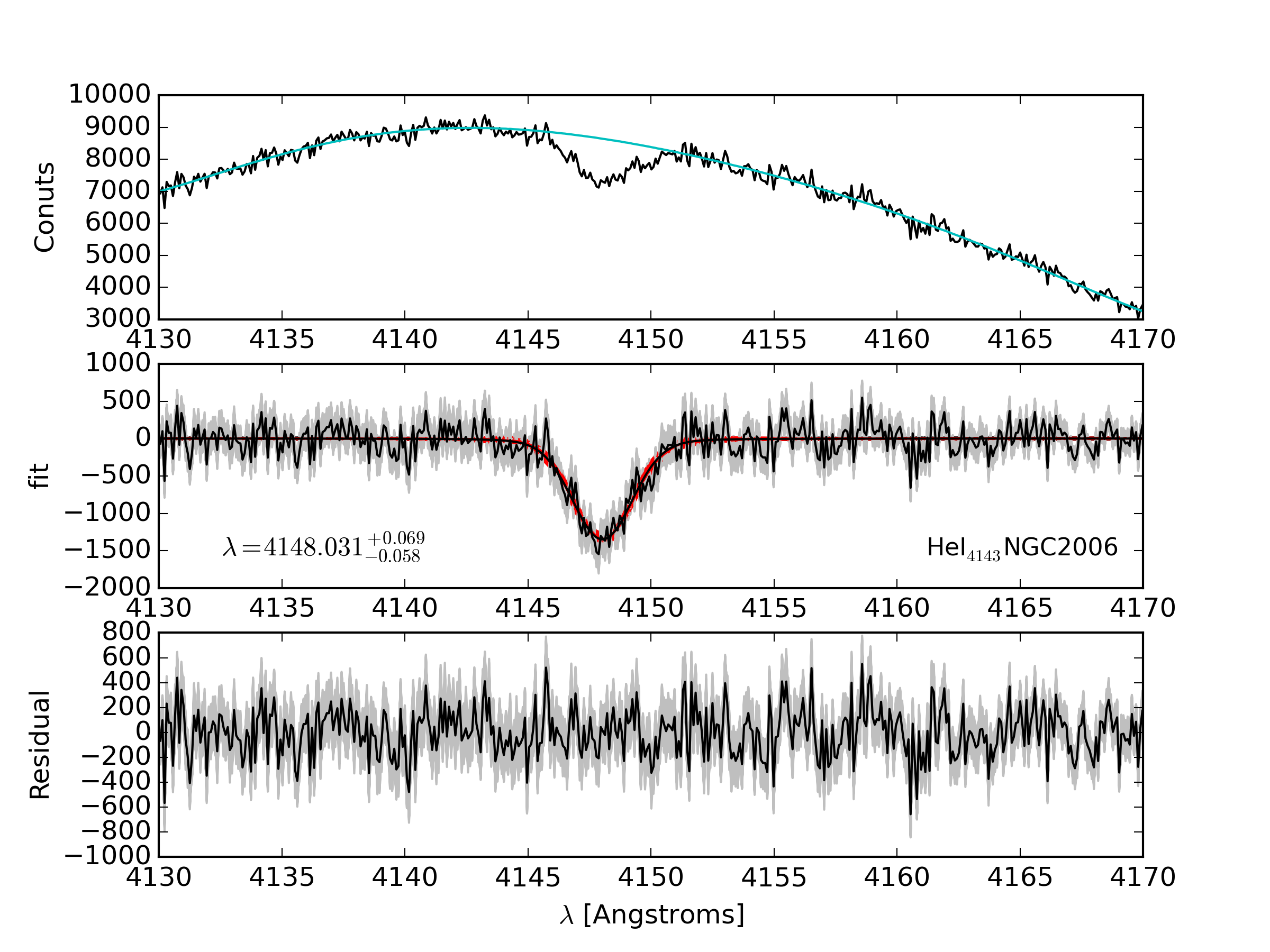}
\includegraphics[width=0.49\textwidth,angle=0]{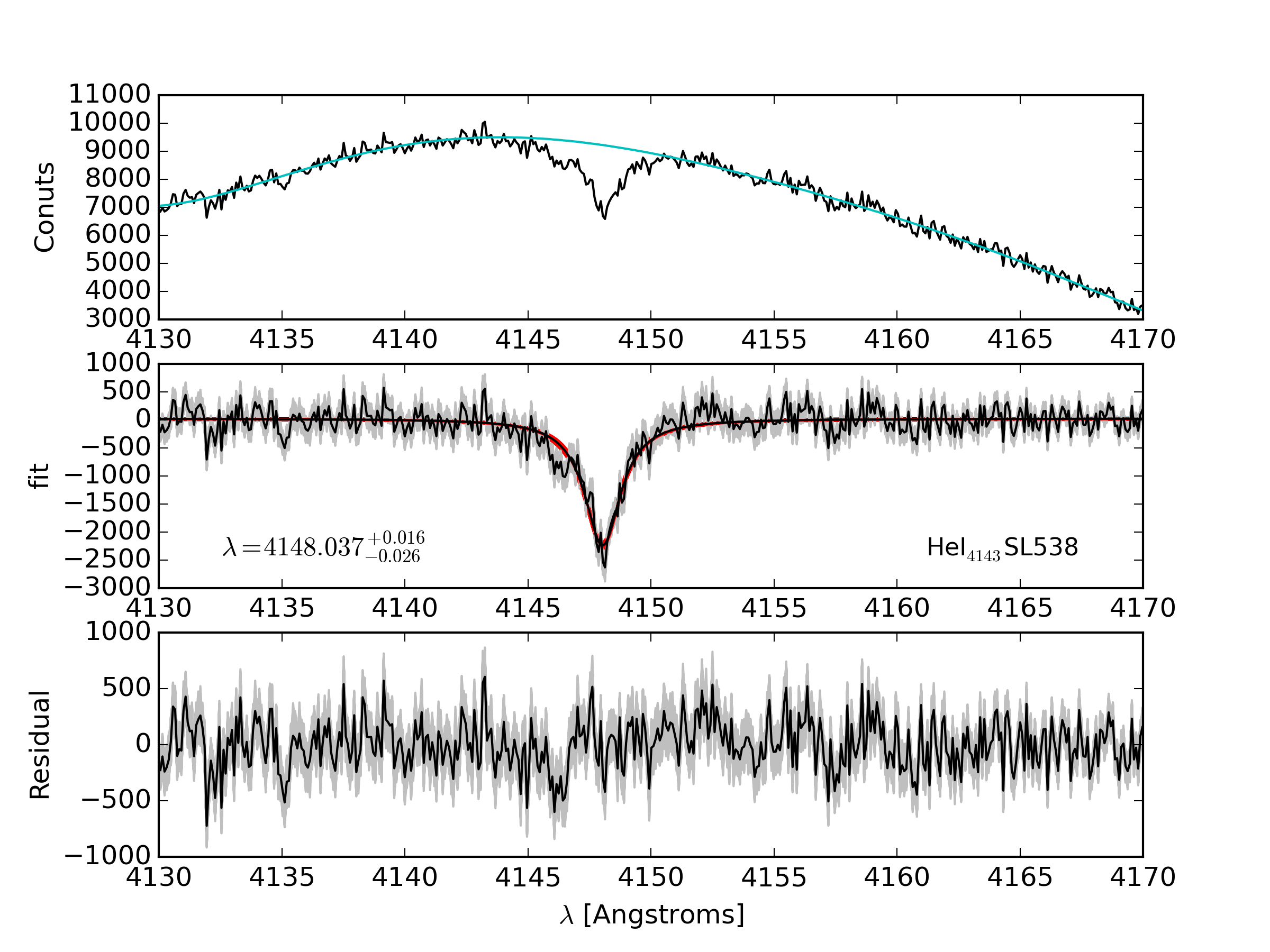}
\caption{-- continued. This figure shows profile fits for the absorption features He\,{\sc i}\,(4009.256\AA), H$_{\delta}$, and He\,{\sc i}\,(4143.761\AA).}
\end{figure*}

\addtocounter{figure}{-1}
\begin{figure*}
\centering
\includegraphics[width=0.49\textwidth,angle=0]{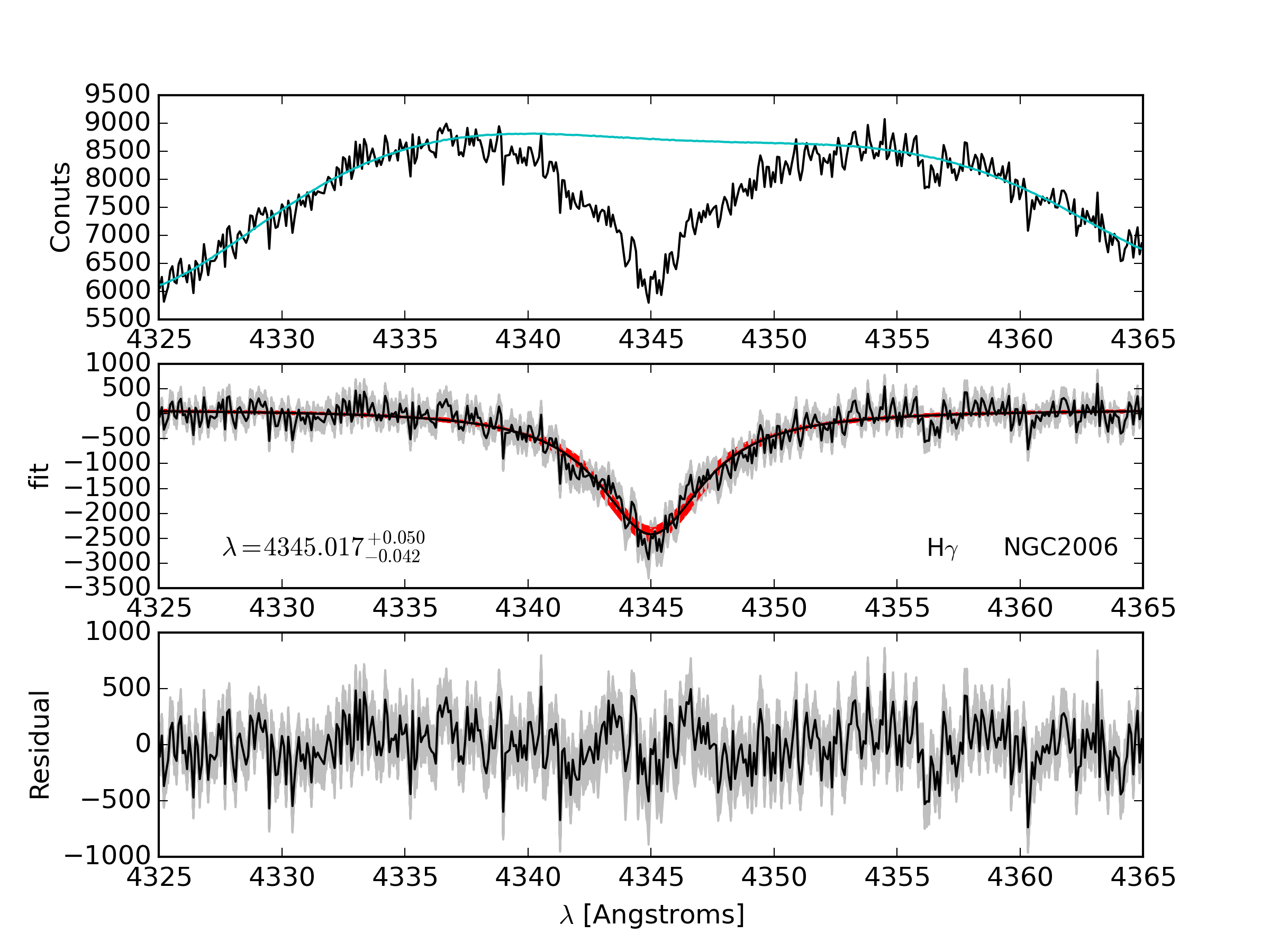}
\includegraphics[width=0.49\textwidth,angle=0]{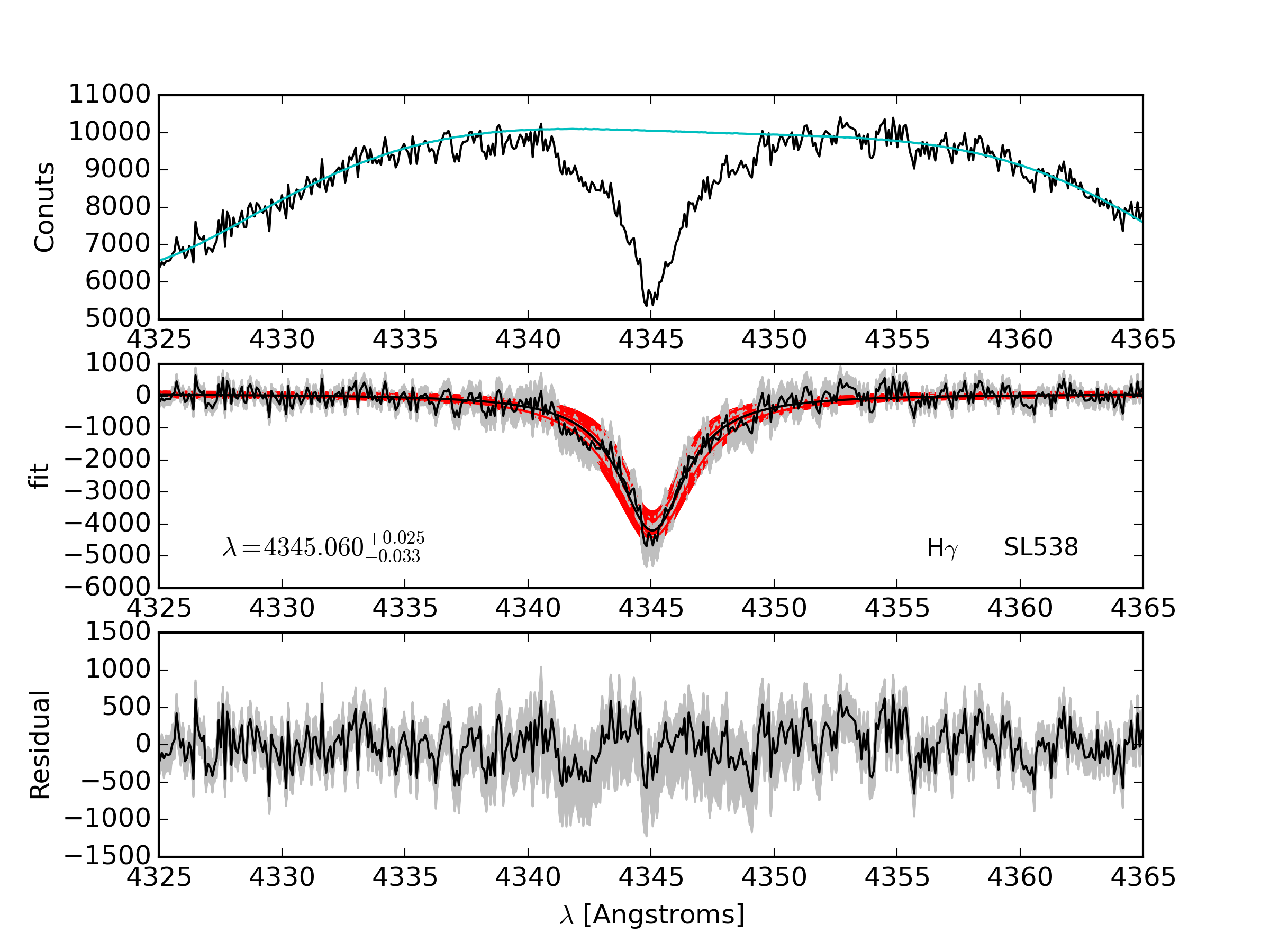}
\includegraphics[width=0.49\textwidth,angle=0]{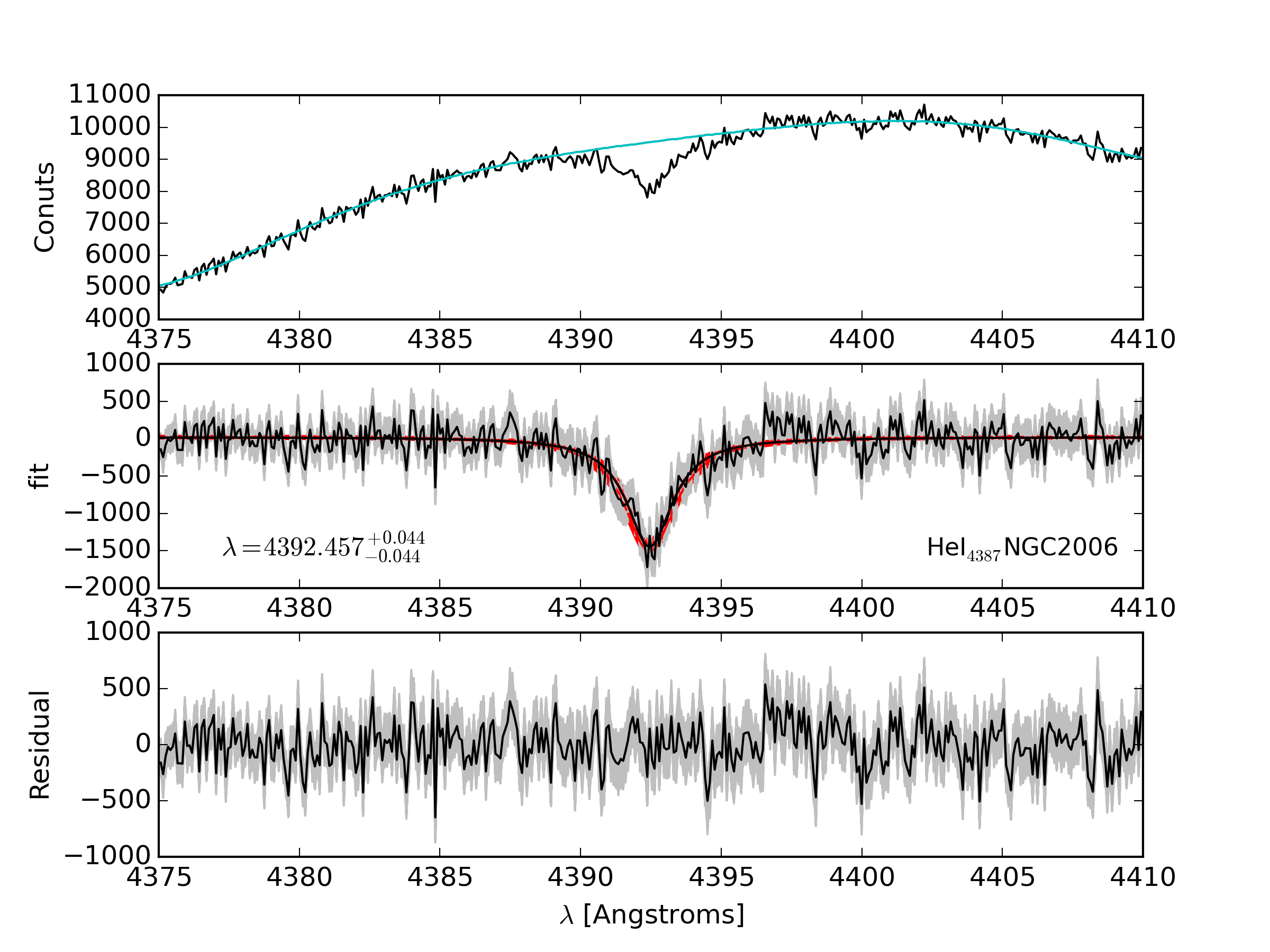} 
\includegraphics[width=0.49\textwidth,angle=0]{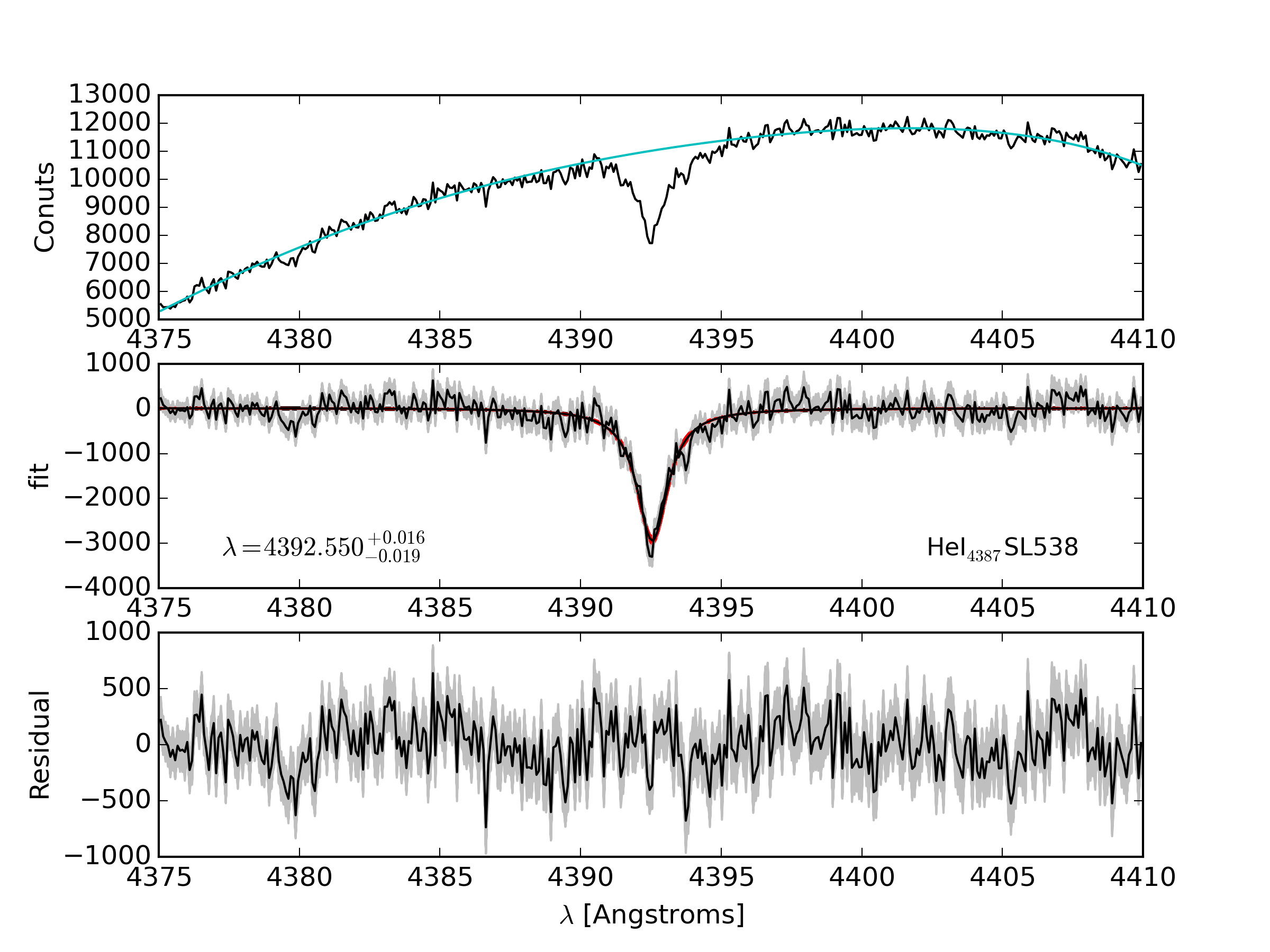}
\includegraphics[width=0.49\textwidth,angle=0]{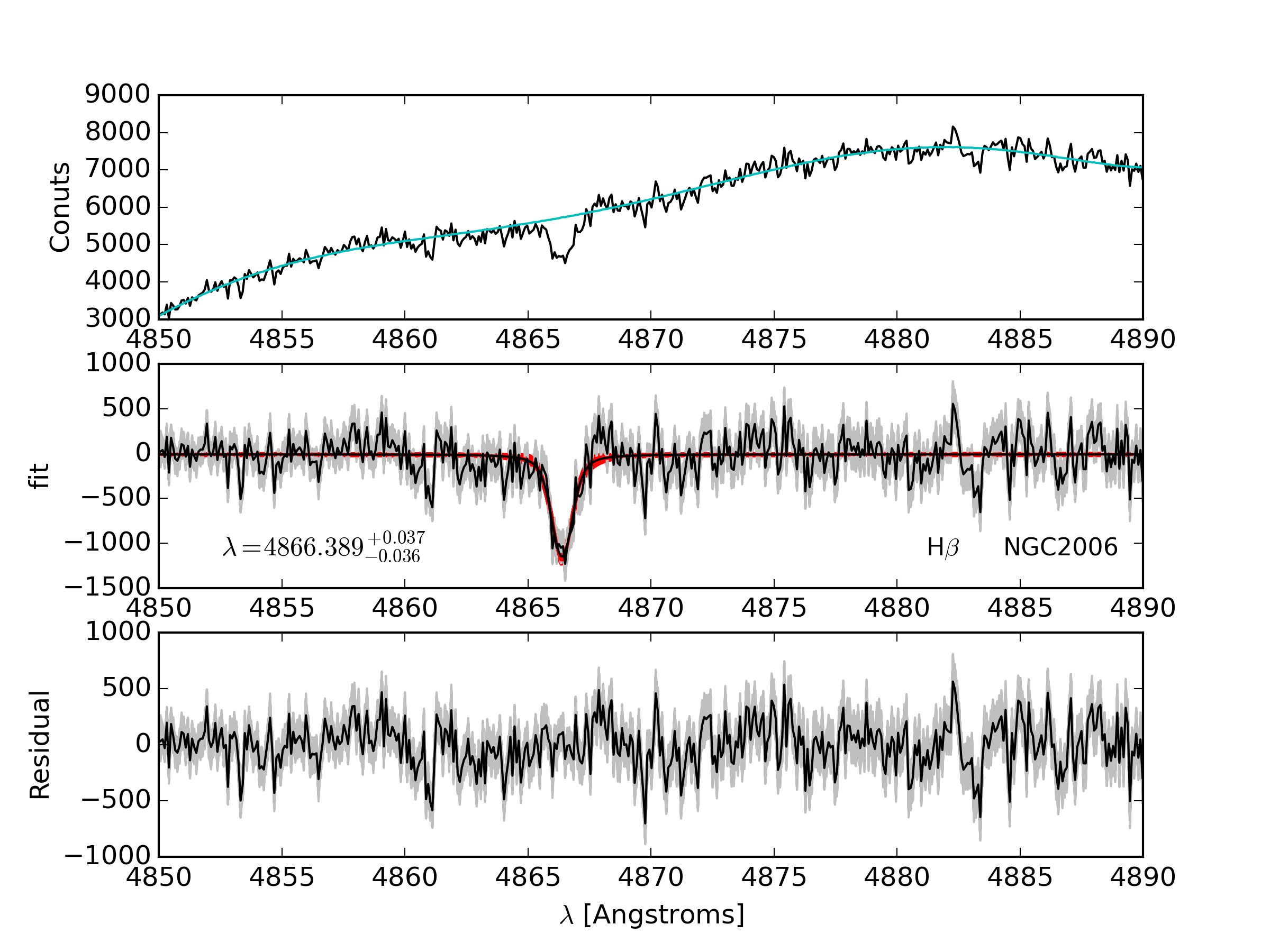}
\includegraphics[width=0.49\textwidth,angle=0]{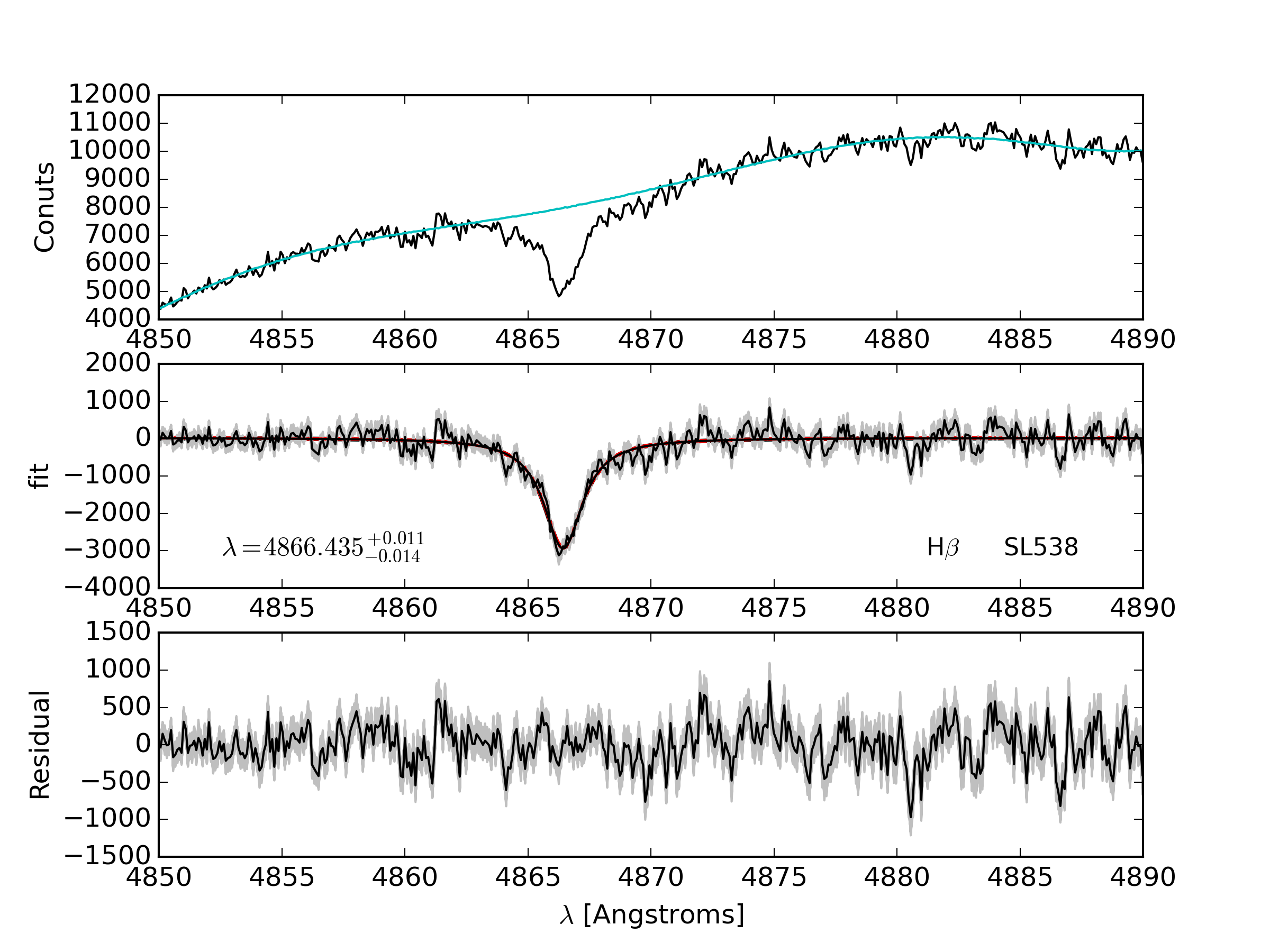}
\caption{-- continued. This figure shows profile fits for the absorption features H$_{\gamma}$, He\,{\sc i}\,(4387.929\AA), and H${_\beta}$.}
\end{figure*}

\section{Absorption feature comparisons}
\begin{figure*}[!ht]
\centering
\includegraphics[width=0.49\textwidth,angle=0]{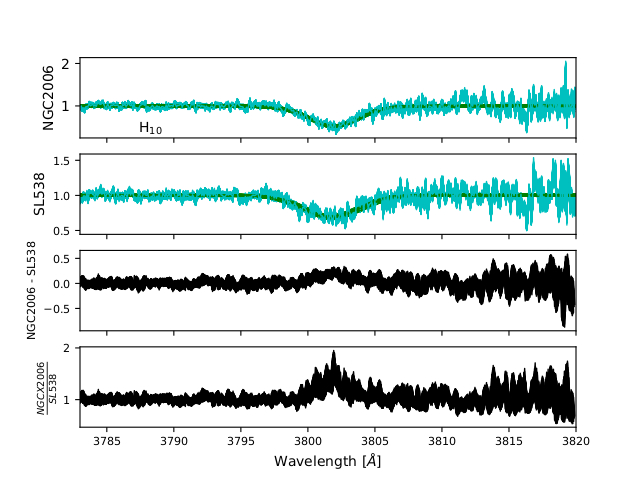}
\includegraphics[width=0.49\textwidth,angle=0]{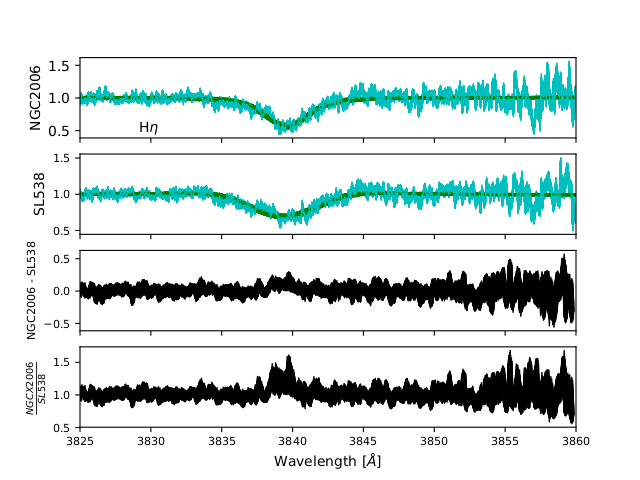}
\includegraphics[width=0.49\textwidth,angle=0]{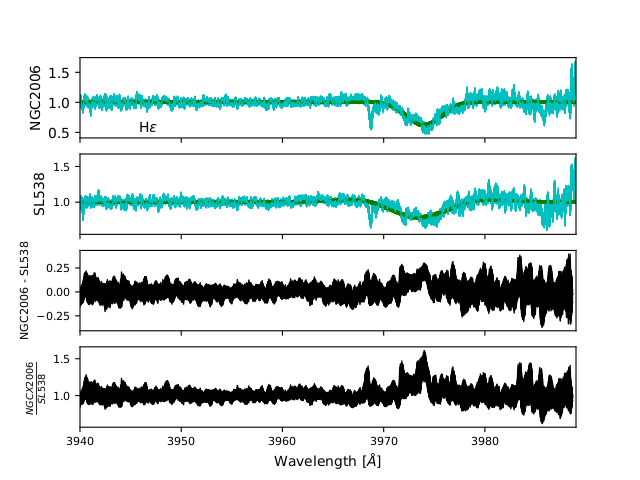}
\includegraphics[width=0.49\textwidth,angle=0]{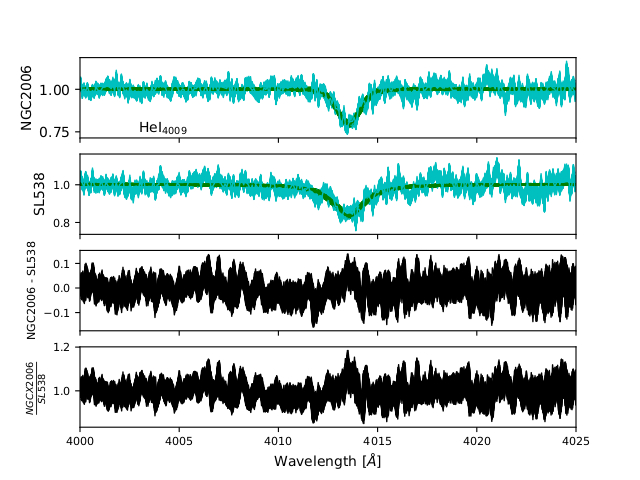}
\includegraphics[width=0.49\textwidth,angle=0]{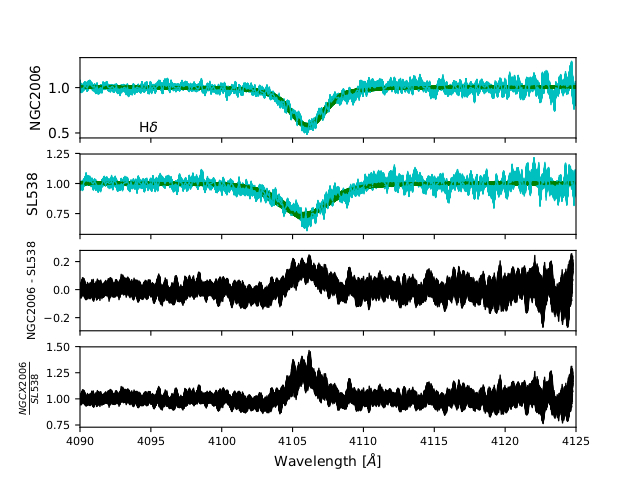}
\includegraphics[width=0.49\textwidth,angle=0]{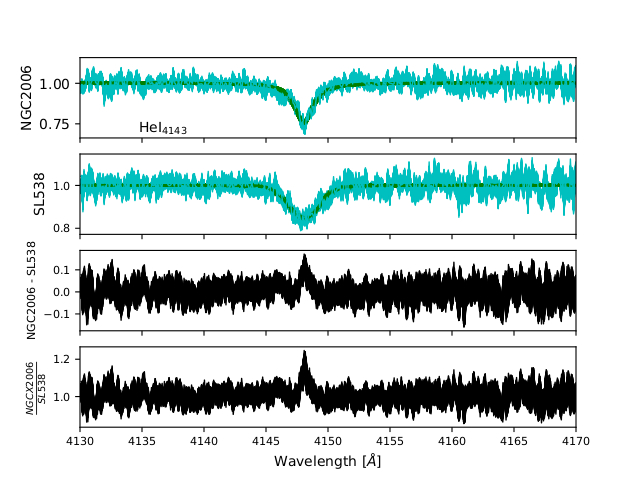}
\caption{Absorption features comparison for H$_{10}$, H$_{\eta}$, H$_{\epsilon}$, He\,{\sc i}\,(4009.256\AA), H$_{\delta}$, and He\,{\sc i}\,(4143.761\AA). Each pair of spectra has been bootstrapped and radial-velocity corrected to be in the same rest-frame. The top panel of each figure shows the absorption feature spectrum portion for NGC\,2006, while the lower panel shows the corresponding section for SL\,538; the next panel below shows the normalized flux difference spectrum (NGC\,2006 -- SL538) and the bottom panel illustrates the flux ratio spectrum (NGC\,2006/SL538). The increasing noise seen on the right in some of the panels corresponds to the decreasing S/N of the spectra.}
\label{Comparison}
\end{figure*}

\addtocounter{figure}{-1}
\begin{figure*}
\centering
\includegraphics[width=0.49\textwidth,angle=0]{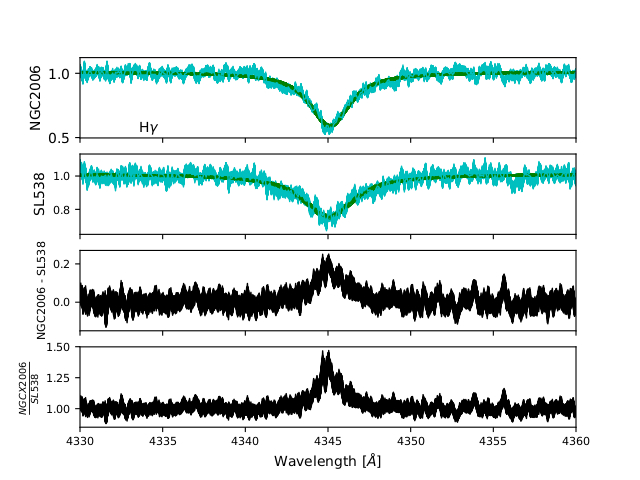}
\includegraphics[width=0.49\textwidth,angle=0]{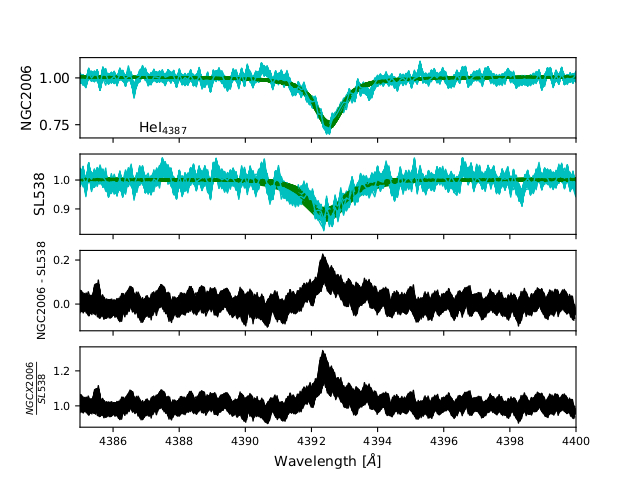}
\includegraphics[width=0.49\textwidth,angle=0]{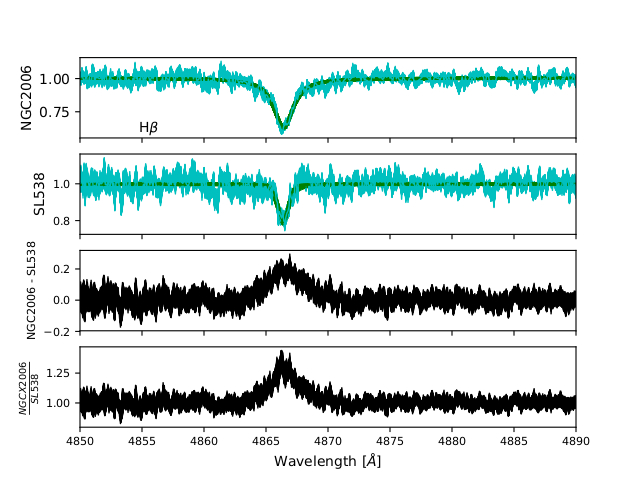}
\caption{-- continued. This figure shows profile comparison for H$_{\gamma}$, He\,{\sc i}\,(4387.929\AA), and H${_\beta}$.}
\end{figure*}

\end{appendix}

\end{document}